\documentclass[twocolumn]{openjournal}

\usepackage{graphicx}
\usepackage{booktabs}
\usepackage{multirow}
\usepackage{tabularx}
\usepackage{placeins}
\usepackage{comment}
\usepackage[dvipsnames]{xcolor} 
\usepackage{lineno}
\usepackage[sort&compress]{natbib}
\usepackage[colorlinks=true, linkcolor=blue, citecolor=blue, urlcolor=blue]{hyperref}
\usepackage[figuresright]{rotating}

\begin{document}

\title{An archival summary: 15 years of ALMA observations on \\disks and planet formation}

\author{Nicol\'as T. Kurtovic$^{1}$, Lizxandra Flores-Rivera$^{1}$, Laura M. P\'erez$^{2}$, Miguel Vioque$^{3}$, Myriam Benisty$^{1}$, Felipe Alarc\'on$^{4}$, Marcelo Barraza-Alfaro$^{5}$, Pietro Curone$^{2}$, Kiyoaki Doi$^{1}$, Sierra Grant$^{6}$, Haochang Jiang$^{1}$, Akimasa Kataoka$^{7}$, Feng Long$^{8}$, \'Alvaro Ribas$^{9}$, Anibal Sierra$^{10}$, Lucas Stapper$^{1}$, Milou Temmink$^{11}$, Francesco Zagar\'ia$^{1}$}

\affiliation{$^{1}$ Max-Planck-Institut für Astronomie, Königstuhl 17, 69117 Heidelberg, Germany}
\affiliation{$^{2}$ Departamento de Astronom\'ia, Universidad de Chile, Camino El Observatorio 1515, Las Condes, Santiago, Chile}
\affiliation{$^{3}$ European Southern Observatory, Karl-Schwarzschild-Str. 2, 85748 Garching bei M\"unchen, Germany}
\affiliation{$^{4}$ Dipartimento di Fisica, Universit\'a degli Studi di Milano, Via Celoria 16, 20133 Milano, Italy}
\affiliation{$^{5}$ Department of Earth, Atmospheric, and Planetary Sciences, Massachusetts Institute of Technology, Cambridge, MA 02139, USA}
\affiliation{$^{6}$ Earth and Planets Laboratory, Carnegie Institution for Science, 5241 Broad Branch Road, NW, Washington, DC 20015, USA}
\affiliation{$^{7}$ National Astronomical Observatory of Japan, Osawa 2-21-1, Mitaka, Tokyo 181-8588, Japan}
\affiliation{$^{8}$ Department of Astronomy, School of Physics, Peking University, Beijing 100871, People's Republic of China}
\affiliation{$^{9}$ Institute of Astronomy, University of Cambridge, Madingley Road, Cambridge CB3 0HA, UK}
\affiliation{$^{10}$ Universidad Nacional Aut\'onoma de México. Instituto de Astronom\'ia. A.P. 70-264, 04510. Ciudad de M\'exico, M\'exico.}
\affiliation{$^{11}$ Leiden Observatory, Leiden University, PO Box 9513, 2300 RA Leiden, The Netherlands}


\begin{abstract}
The Atacama Large (sub-)millimeter Array (ALMA) has been in scientific operations for almost 15 years. We celebrate this achievement by providing a summary of the ``Disks and planet formation'' scientific category, with an emphasis on the disks located in the nearby star-forming regions. As of the beginning of February 2026, ALMA had observed 3933 independent coordinates, which we analyzed by their location in the sky, frequency coverage, exposure time, spectral line coverage, and angular resolution. 
We encourage the community to explore new scientific questions that are made possible through the archival datasets. 
\end{abstract}

\maketitle

\section{Introduction}

The unique technical capabilities of the Atacama Large (sub-)Millimeter Array (ALMA) have allowed studies of disks and planet formation with an unprecedented combination of angular resolution and sensitivity at millimeter wavelengths. In the planet formation community, ALMA has allowed surveys of large numbers of objects for population studies \citep[e.g.,][]{ansdell2016, cieza2019, pascucci2016, vanterwisga2022}, observing substructures in the dust continuum emission at very high angular resolution \citep[e.g.,][]{almapartnership2015, andrews2016, andrews2018b, andrews2020, bae2023, guerra-alvarado2025, huang2018b, long2018b}, molecular line inventories for astrochemistry \citep[e.g.,][]{cleeves2018, miotello2017, miotello2019, oberg2021b, pegues2020, zhang2025}, as well as studies of gas kinematics \citep[e.g.,][]{casassus2019, izquierdo2022, pinte2019, pinte2023, teague2018a, teague2019, teague2025a}. These observations provide insights into planet forming disk properties, allowing for constraints of the disk's solid and gaseous mass reservoirs, composition, structure, temperature, and radial extent. 

The growing volume of archival data is enabling research projects that require the combination of multiple observations, as well as the possibility of revisiting targets with updated analysis techniques. This is demonstrated by the steady increase in publications based partially or completely on ALMA archival data, which reached 40\% of all ALMA related publications in 2024 \citep{stoehr2026}. In practice, the ALMA archive has become a large survey of planet forming disks, where hundreds of different programs contribute to a heterogeneous sample in sensitivity, frequency coverage, angular resolution, and distribution of time among targets. 

As of April 2026, ALMA has been observing disks for almost 15 years, and understanding the contents of the available observations is increasingly relevant for archival mining and proposal planning. In this article, we aim to provide a summary of the accumulated archival data over this decade and a half, with a focus on the scientific category of ``Disks and planet formation''. In Sect.~\ref{sec:methodology}, we describe our methods for identifying independent sky coordinates in the archive, and our approach at cross-matching these coordinates with the literature. In Sect.~\ref{sec:results}, we present a summary of the archival data organized by coordinates, ALMA Band, time on source, star-forming region, ALMA project code, spectral line coverage, and angular resolution. We discuss the distribution of ALMA time and publication efficiency in Sect.~\ref{sec:discussion}. Finally, our main conclusions are summarized in Sect.~\ref{sec:conclusions}.

\begin{figure*}
    \centering
    \includegraphics[width=0.96\textwidth]{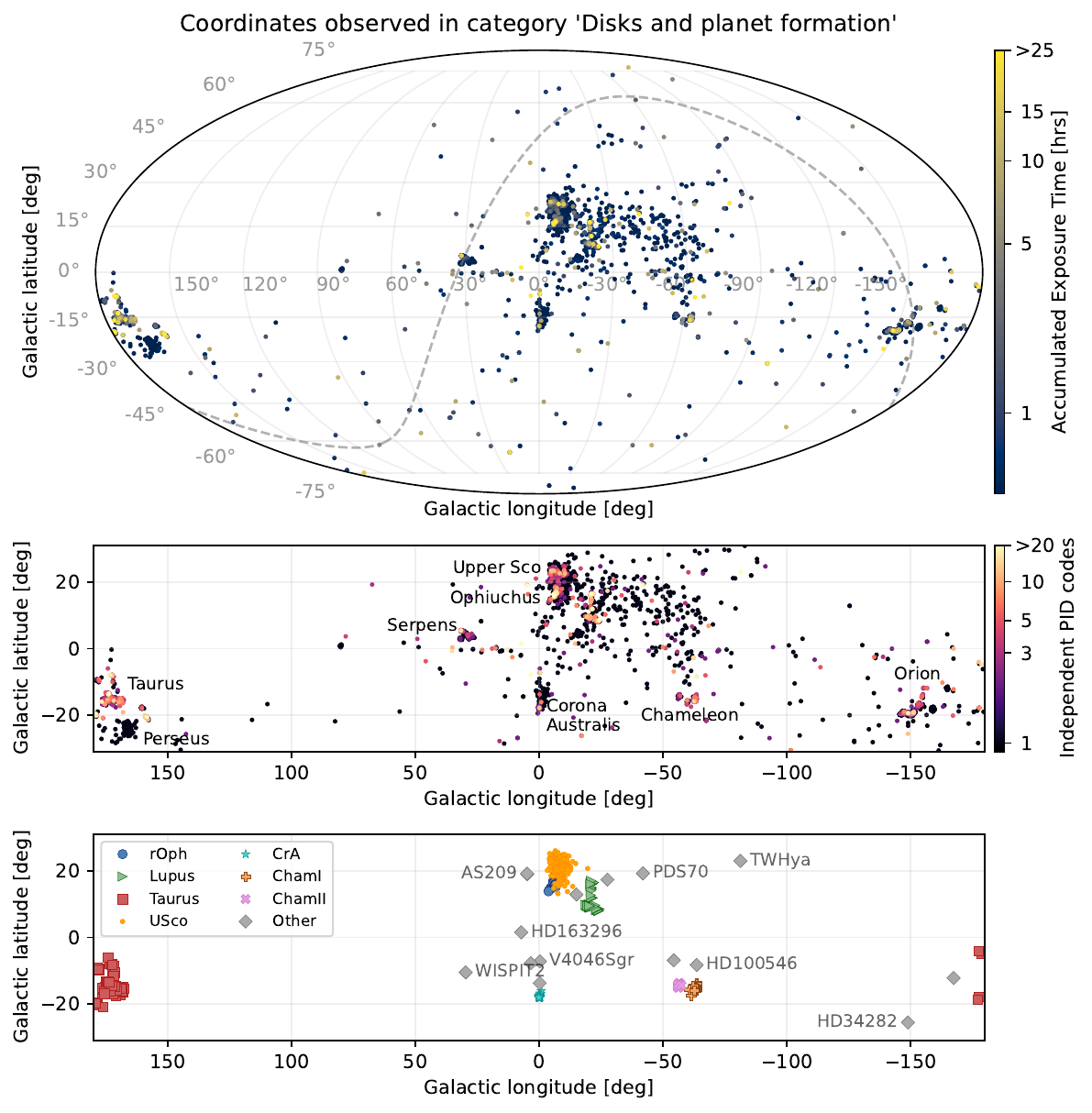}
    \caption{Distribution of coordinates observed as part of the scientific category ``Disks and planet formation''. In the top panel, the color scaling shows the accumulated exposure time per coordinate, including all ALMA Bands and time on source from other categories. The dashed line shows the coordinates where the Declination is 0, separating northern from southern sky. In the central panel there is a cut out of the galactic plane, with the color scale showing the number of independent project codes that have observed each coordinate. In the bottom panel, we show the matching  region from the extended PPVII table.}
    \vspace{4mm}
    \label{fig:sky-comb-coordinates}
\end{figure*}

\begin{figure*}
    \centering
    \includegraphics[width=0.95\textwidth]{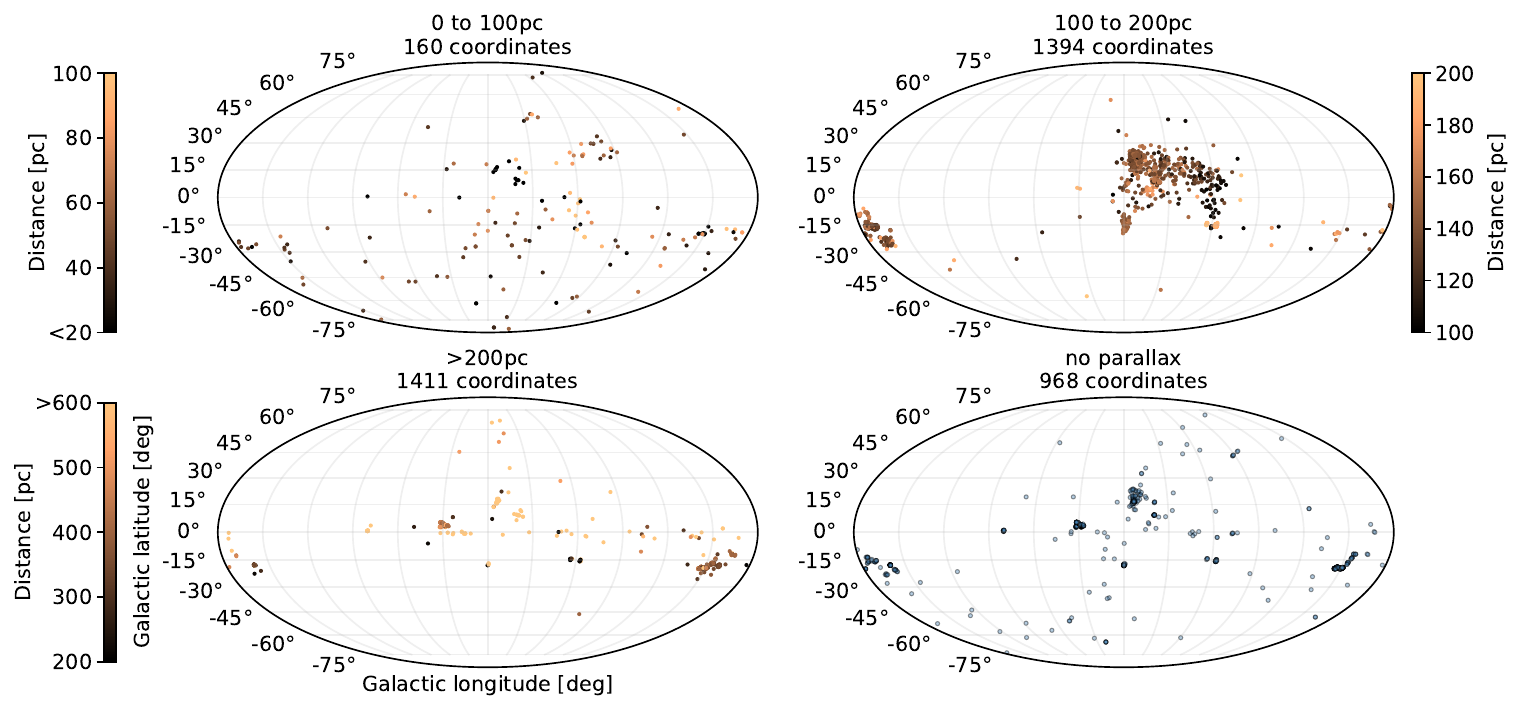}
    \caption{Distribution of coordinates observed as part of the scientific category ``Disks and planet formation'', grouped by their estimated distance from Gaia DR3 parallaxes. Note that each panel has its own colorbar, with the exception of the disks with no parallax from Gaia. }
    \vspace{4mm}
    \label{fig:sky-distance-all}
\end{figure*}

\section{Methodology}\label{sec:methodology}

\subsection{Identifying independent coordinates}

In order to find all observations associated to the ``Disks and planet formation'' category, we downloaded the full list of observations from the web interface of the European ALMA Archive\footnote{\url{https://almascience.eso.org/aq/}} on February 2nd 2026, by exporting the table of ``Observations'' in \texttt{CSV} format after filtering for data obtained with the 12\,m array. The table entries include all observations taken since the beginning of scientific operations, for which we recover their exposure time on source, frequency coverage, average angular resolution, ALMA Band, project code, and scientific category. We note that the total 12\,m time (including overheads from calibrations) is not publicly available through the web interface, and thus we only compare observations based on their exposure time on source. 

The coordinates of a specific science target might change over the years due to their proper motion, or due to small differences in coordinate reference systems. Therefore, a threshold tolerance in sky distance is needed to avoid duplications within the list of observations. In order to identify independent coordinates observed by ALMA, we set a maximum distance of $5''$ between phase centers to be considered as the same target. This larger threshold also avoids duplications for binary sources that fall within the same field of view of their companions, which in this work will be registered as a single coordinate. Thus, in the following sections, we refer to ``coordinates'' instead of ``targets'', as the same coordinate could contain observations for more than one source (e.g., binary systems). 

After compiling all the independent coordinates, we searched for all associated observations to each of them, regardless of their original scientific category. For example, the coordinate associated to ``HL Tau'' has been observed in the ``Disks and planet formation category'', but it also has been targeted under ``Solar system'' and ``ISM and star formation''. Thus, a full accounting of the time spent on HL\,Tau needs to include observations from all relevant ALMA project codes across categories. The same procedure is applied to every independent coordinate, for which we record the number of distinct project codes and total exposure time in each ALMA Band. We complete our table by querying the Gaia DR3 archive to obtain the closest parallax measurement within $5''$, and estimate the distance to the science target in the coordinate as the inverse of the parallax.

\subsection{Matching observed coordinates with the literature}

Our list of independent ALMA coordinates was matched to the PPVII table\footnote{\url{https://ppvii.org/chapter/15/}} of \citet{manara2023}, which compiles a list of planet forming disks in nearby star-forming regions, containing 845 targets distributed across Ophiuchus, Lupus, Taurus, Upper Sco, Corona Australis, Chameleon I, and Chameleon II \citep[compiling millimeter properties from ][we refer the reader to \citet{manara2023} for a detailed description of the sample construction]{akeson2014, akeson2019, andrews2013, ansdell2016, ansdell2018, barenfeld2016, carpenter2014, cazzoletti2019, cieza2019, long2018a, long2019, pascucci2016, sanchis2020, vanderplas2016, villenave2021, ward-duong2018, williams2019}. We complemented this list with the missing targets from the published ALMA Large Programs (DSHARP, \citet{andrews2018b}; MAPS, \citet{oberg2021b}; AGE-PRO, \citet{zhang2025}; exoALMA, \citet{teague2025a}), from the latest ALMA survey of Upper Sco \citep{carpenter2025a}, as well as additional notable sources found among those with the longest accumulated exposure times with ALMA (see Sect.~\ref{sec:sec:accumulated-exposure-time}). We refer to this table with additional targets as the extended PPVII table. 


Whenever possible, the new targets were included as part of their  region (e.g., HL\,Tau in Taurus, HD\,142666 in Upper Sco). If the target is not associated to any of the original PPVII regions, we included it in a category of ``Other'' (e.g., TW\,Hya, V4046\,Sgr, WISPIT\,2).
We iterated over our list of independent ALMA coordinates to find the nearest target in the extended PPVII table, with a maximum search radius of $5''$. When a match is found, we associate the independent coordinate to the name of the source, the  region, and stellar mass. We note that this table is only for nearby star-forming regions, and thus a dedicated analysis including Orion, Serpens, or debris disks, is not included in this work.

\begin{figure*}
    \centering
    \includegraphics[width=0.99\textwidth]{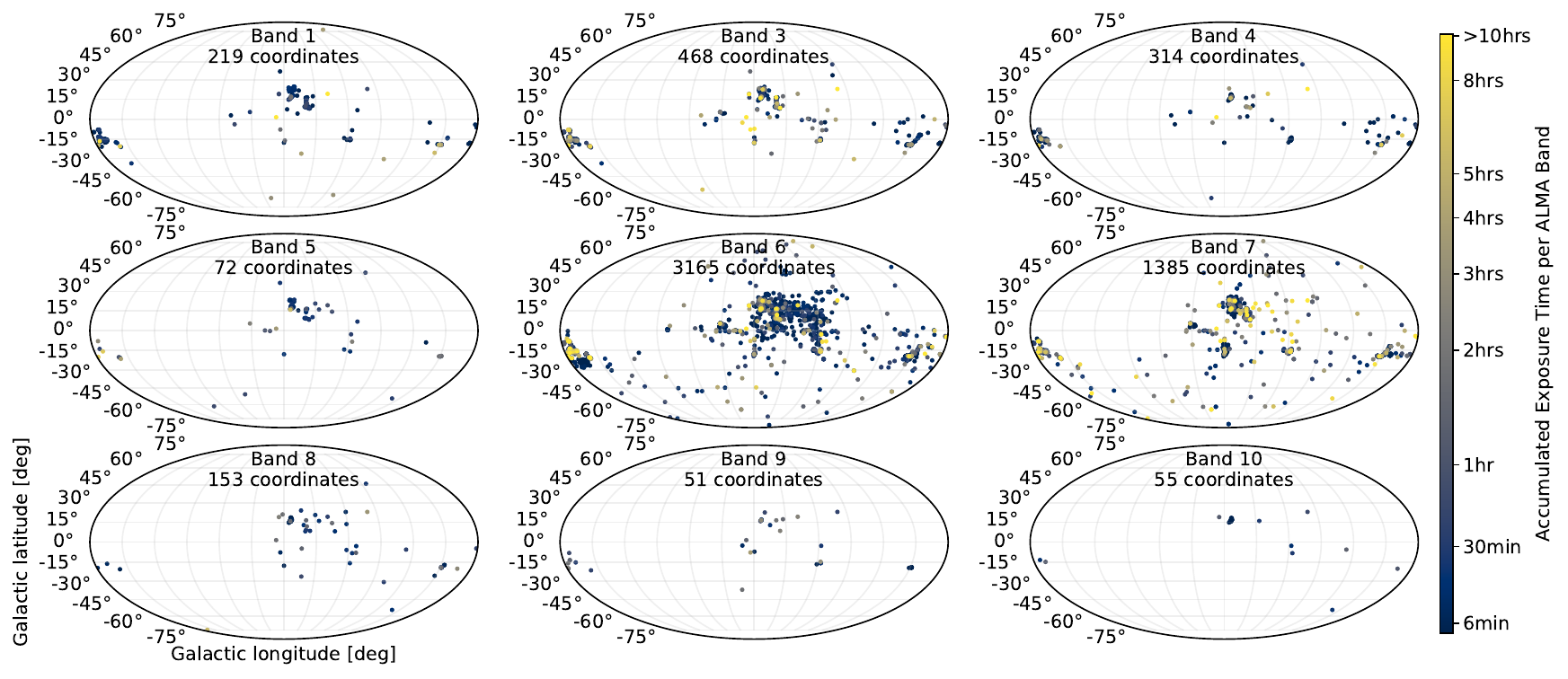}
    \caption{Distribution of coordinates per ALMA Band, observed as part of the scientific category ``Disks and planet formation''. The color scale shows the accumulated time in each specific Band, for each coordinate. }
    \vspace{4mm}
    \label{fig:sky-bandtime}
\end{figure*}

\subsection{Information per ALMA Project code}

In addition to the list of observations, we also downloaded the list of ALMA project codes, spanning all observed projects from ALMA Cycle 0 to the first third of ALMA Cycle 12, including the project code, title and number of publications. We cross matched the table of observations to recover the number of independent coordinates included in each project code. Similarly, we estimate the total exposure time of a project code by integrating all its individual entries in the archive. In the following sections, the results relative to project codes are always shown only for those in the scientific category of ``Disks and planet formation''.

\bigskip

\section{Results}\label{sec:results}

\subsection{By coordinates}

We identified a total of 3933 independent coordinates observed by ALMA as part of the ``Disks and planet formation'' category, spanning from Cycle 0 to the beginning of Cycle 12 (as of February 2026, ALMA had completed four months of Cycle 12). These coordinates are presented in Fig.~\ref{fig:sky-comb-coordinates}. The majority of the coordinates are not randomly distributed over the sky, as clusters can be identified around the known star-forming regions, as illustrated in the middle and lower panel of Fig.~\ref{fig:sky-comb-coordinates}. 

The distribution of these coordinates becomes clearer when separating them by estimated distance, as presented in Fig.~\ref{fig:sky-distance-all} (also shown in in the galactic plane in Fig.~\ref{fig:app:galactic_plane} in the Appendix). From 0\,pc to 100\,pc, we find 160 coordinates distributed across the sky. While some notable planet forming disks are included among these coordinates (TW\,Hya, V4046\,Sgr, MP\,Mus), the spatial spread in their distribution is explained by the projects targeting debris disks, or searching for millimeter emission of nearby stars (e.g., TRAPPIST-1, \citealt{hughes2019, marino2020}; Proxima Cen, \citealt{anglada2017, burton2025}).

\begin{figure*}
    \centering
    \includegraphics[width=0.99\textwidth]{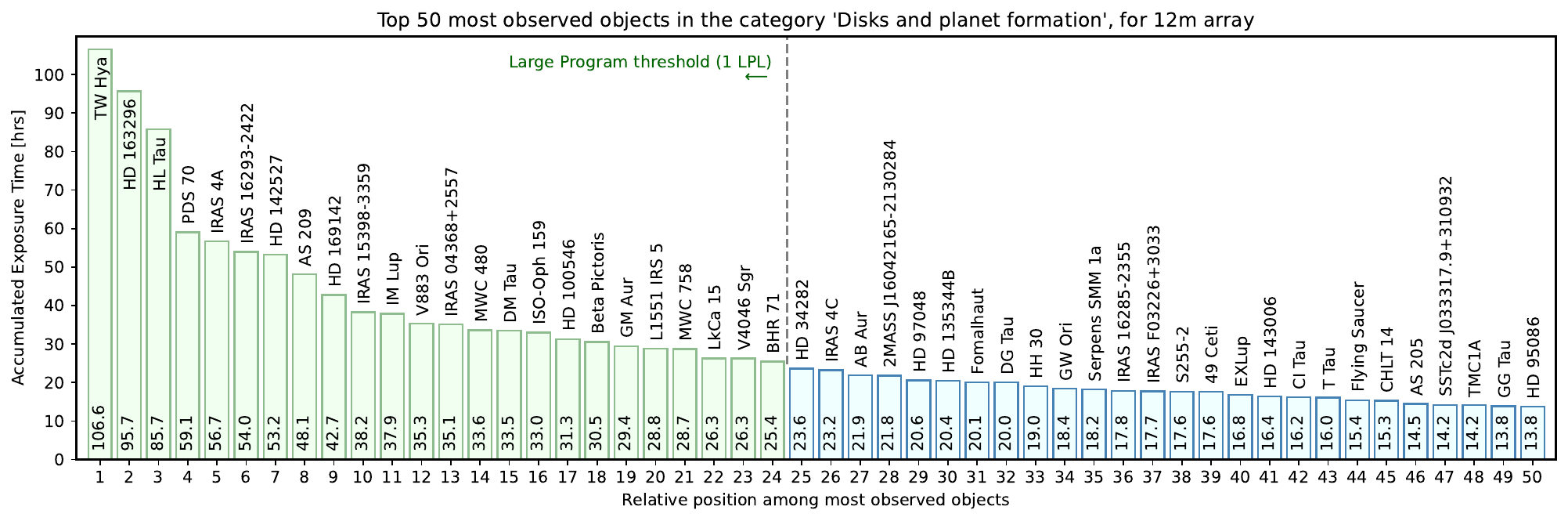}
    \caption{Sources with the longest accumulated exposure time in the ALMA Archive, considering all observations taken with the 12\,m array since ALMA Cycle 0. While the sources were selected among those in the ``Disks and planet formation'' science category, here we integrate the exposure time across observations from any science category (e.g., from ``ISM and star formation'' if available). In green, we show the sources with more exposure time than the shortest Large Program \citep[2016.1.00484.L,][]{andrews2018b, huang2018b}.}
    \vspace{4mm}
    \label{fig:hist-timesource}
\end{figure*}

Between 100\,pc and 200\,pc, two large clusters of coordinates are distinguished. Near the center of the galactic coordinates we identify the large Sco-Cen association \citep[see][]{ratzenbock2023}, which spans from Corona Australis, extending through Ophiuchus, Upper Sco, Lupus, and wrapping toward Chameleon I and II (see location of star-forming regions in Fig.~\ref{fig:sky-comb-coordinates}). On the opposite side of the sky we see the clusters containing the regions of Taurus and Perseus. Beyond 200\,pc, we identify coordinates in the regions of Chameleon I and II, Serpens, Orion, as well as sources along the galactic plane.

A total of 968 coordinates are not associated to a parallax from Gaia DR3 in their surrounding $5''$, potentially due to their optical faintness. Nevertheless, the clustering of coordinates missing parallax values suggests that many of these sources are part of regions like Ophiuchus, Serpens, Orion, Chameleon, and Taurus.

From Fig.~\ref{fig:sky-comb-coordinates}, we can also identify the coordinates that have been observed by several ALMA project codes (i.e., several independent accepted ALMA proposals). In the middle panel of Fig.~\ref{fig:sky-comb-coordinates}, brighter colors show which coordinates were more often observed in different ALMA projects, roughly coinciding with the nearby star-forming regions, as well as Orion. Considering that ALMA projects are discouraged from duplicating existing observations, coordinates with multiple projects can be associated to a more diverse set of available data, either in sensitivity, angular resolution, or wavelength.

The distribution and number of coordinates observed in each ALMA Band is presented in Fig.~\ref{fig:sky-bandtime}. Most coordinates have at least one observation in Band 6, with 3165 coordinates ($80.5\%$ of the total), followed by Band 7 with 1385 ($35.2\%$ of the total). The longest Band 6 exposures are concentrated in nearby star-forming regions, whereas the longest Band 7 integrations are more spatially distributed (likely from observations of project codes 2021.1.01123.L and 2022.1.00338.L). Band 3 is the third most common band in number of coordinates, with clusters coinciding with Ophiuchus, Lupus, Taurus, and Orion. Band 1, which has only been available in Cycle 11 and 12, is already the fourth Band with the highest number of coordinates. 

In contrast, measurements in the high-frequency Bands 9 and 10 have only been obtained for 51 and 55 coordinates, respectively, representing about $1.4\%$ of the total. In Band 10, which has been available since Cycle 3, the majority of observed coordinates correspond to project 2023.1.00131.S, a Cycle 10 snapshot survey of Ophiuchus with 31\,s per source. The remaining Band 10 targets include a small number of Class II disks (AS\,209, TW\,Hya, RY\,Tau\footnote{RY\,Tau and HD\,100546 were observed in Band 10 as part of the ``ISM and star formation'' scientific category.}, HD\,100546) and debris disks (among them, HD\,131488, HD\,21997, HD\,76582, HD\,61005). Excluding debris disks, the accumulated Band 10 time dedicated to planet-forming disks is 4\,hrs on source in total.

\subsection{By accumulated exposure time per disk}\label{sec:sec:accumulated-exposure-time}

We integrate the exposure time in all ALMA Bands to recover the total accumulated exposure time on source for each independent coordinate. The median exposure time for coordinates is 3.024\,min on source. If we only consider the coordinates with matching entries in the extended PPVII table, we find a median exposure time of 6.048\,min. The average exposure time, instead, deviates by over an order of magnitude from the median, with 2.02\,hrs (121.4\,min) on source for the coordinates in the extended PPVII table. This large difference is explained by the non-uniform distribution of exposure time among the observed coordinates, which we further discuss in Sect.~\ref{sec:sec:dist-time-sources}. 

We sort the coordinates by total exposure time on source, and for the 50 coordinates with the longest accumulated time we find the corresponding science target with the SIMBAD Astronomical Database \citep{simbad2000}. These results are presented in Fig.~\ref{fig:hist-timesource}. 
The most observed source is TW\,Hya, which accumulates over 106\,hrs on source since the beginning of scientific operations of ALMA. This long exposure time is consistent with TW\,Hya being the closest Class II planet forming disk, which makes it an ideal target for high sensitivity dust continuum and gas emission analysis. 
Following in the ranking we identify: disks with protoplanet candidates \citep[HD\,163296, AS\,209,][]{teague2018a,bae2022}, disks with protoplanet detections  \citep[PDS\,70, HD\,169142,][]{keppler2018,hammond2023}, the brightest millimeter disk in the sky \citep[HL\,Tau,][]{almapartnership2015}, and one of the largest and brightest disks with a high-contrast dust asymmetry \citep[HD\,142527,][]{casassus2013}. The top 10 ranking is completed with the younger objects IRAS\,4A (identified in SIMBAD as ``[JCC87] IRAS 4A''), IRAS\,16293-2422, and IRAS\,15398-3359.

\begin{figure*}
    \centering
    \includegraphics[width=1.0\textwidth]{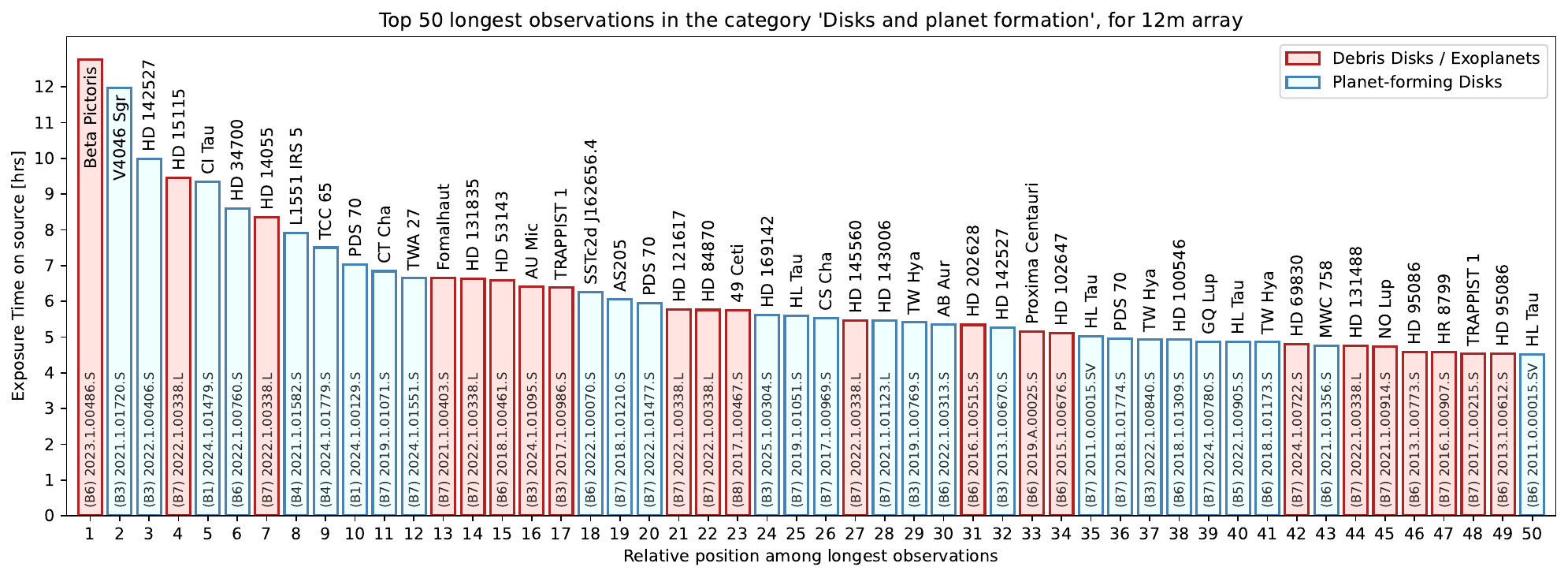}
    \caption{Longest single observation on a single coordinate with a specific frequency setup, excluding existing complementary configuration time (i.e., equivalent to an individual entry in the ALMA archive). Bars are colored depending on their closest relevant scientific category, as shown in the top left legend. The ALMA Band and project code are shown at the bottom of each bar. \\
    \textit{\textbf{Note}: Scientific keywords are set manually by the PI team during the preparation of the proposal. Thus, multiple keywords can be associated with the same object. For example, PDS\,70 has observations with the keywords ``Disks around low-mass stars'', ``Disks around high-mass stars'', and ``Exoplanets''. For the purposes of coloring this figure, we identify all observations of PDS\,70 as being ``planet forming disks''. The same logic is applied to other objects to associate them to their most relevant keyword. }\label{fig:hist-longesttime}}
    \vspace{4mm}
\end{figure*}

For reference, we highlight in Fig.~\ref{fig:hist-timesource} sources with longer exposure time than the shortest Large Program on  disks \citep[DSHARP, 24.4\,hrs on-source][]{andrews2018b}. Although ALMA Large Programs request more than 50\,hrs of the 12\,m array, only about half of that time corresponds to on-source time, with the remaining going into the overheads and calibrations, resulting in an approximated threshold of $\approx24\,$hrs on source, which we define as one Large Program length (LPL).
We identified 24 coordinates that have on-source time at the level of 1 LPL, only considering 12\,m array observations. Among these 24, there are 16 systems typically studied in the context of planet forming disks, 7 systems more commonly associated to studies of star formation, and one debris disk (Beta Pictoris).

\subsection{By longest single observation}\label{sec:sec:longest}

Within the scientific category of ``Disks and planet formation'', we can also explore the longest observations as part of an individual project code, corresponding to a single entry in the archive (thus, a single frequency setup, and not considering complementary baseline observations). The top 50 longest observations are shown in Fig.~\ref{fig:hist-longesttime}, color coded according to their main scientific keyword. Unlike the ranking of accumulated time (Fig.~\ref{fig:hist-timesource}), the ranking of longest individual observation is more populated by debris disks, which require deep integrations due to their faint millimeter continuum emission. 

For each observation in this ranking, we checked the number of associated publications to their corresponding project code. The number of publication counts provided by ALMA are tracked by project code, and not by specific observations. Therefore, separate articles may only use a subset of data from a given project code, while still contributing to increase its publication count. In the context of this archival summary, we interpret the number of publications as a measure for the number of research projects that have been directly enabled by the existence of a dataset. 

Among the observations of Fig.~\ref{fig:hist-longesttime}, we find a difference between the publication count of observations taken as part of community efforts (Large Programs and HL\,Tau observations), and those that come from Regular Programs. As the top 50 longest observations have comparable exposure time (difference is in the order of $\times2$, see Fig.~\ref{fig:hist-longesttime}), we can compare them by the metric of publications per hour (see Sect.~\ref{sec:sec:discussion-pubhr} for a detailed discussion of this metric). We find a $0.22_{-0.05}^{+0.66}$\,pub/hr for Regular programs, and $1.83_{-0.26}^{+3.54}$\,pub/hr for community efforts (Large Programs and HL\,Tau), where error bars are given by the 16th and 84th percentiles of the sample.

\subsection{By  region}\label{sec:sec:by-sfr}

\begin{figure*}
    \centering
    \includegraphics[width=0.75\textwidth]{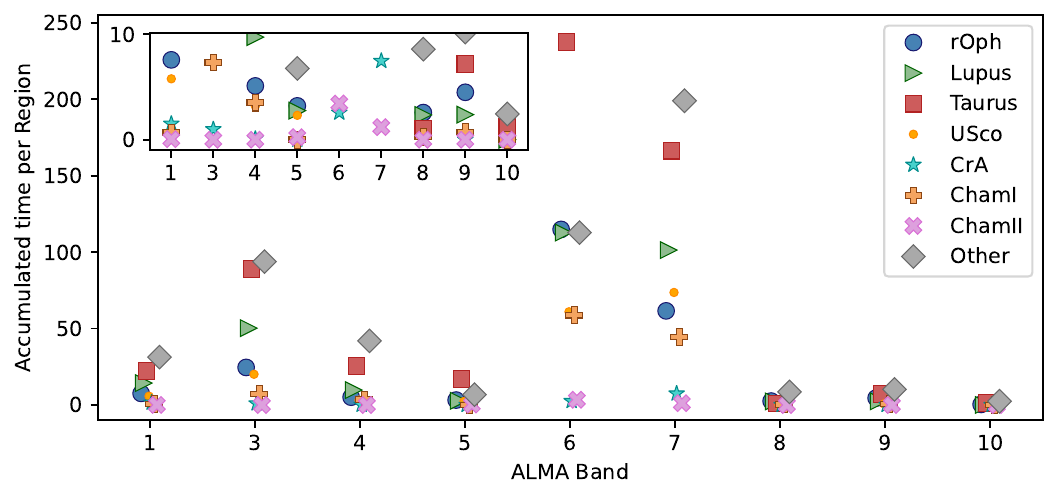}
    \caption{Accumulated exposure time per ALMA Band for the nearby star-forming regions. The inset shows a zoom into the area between 0 and 10\,hrs of accumulated exposure time. Corona Australis and Chameleon II have $\leq2$\,hrs in all Bands different from Band 6 and 7. \label{fig:time-per-region}}
    \vspace{4mm}
\end{figure*}

\begin{figure*}
    \centering
    \includegraphics[width=0.99\textwidth]{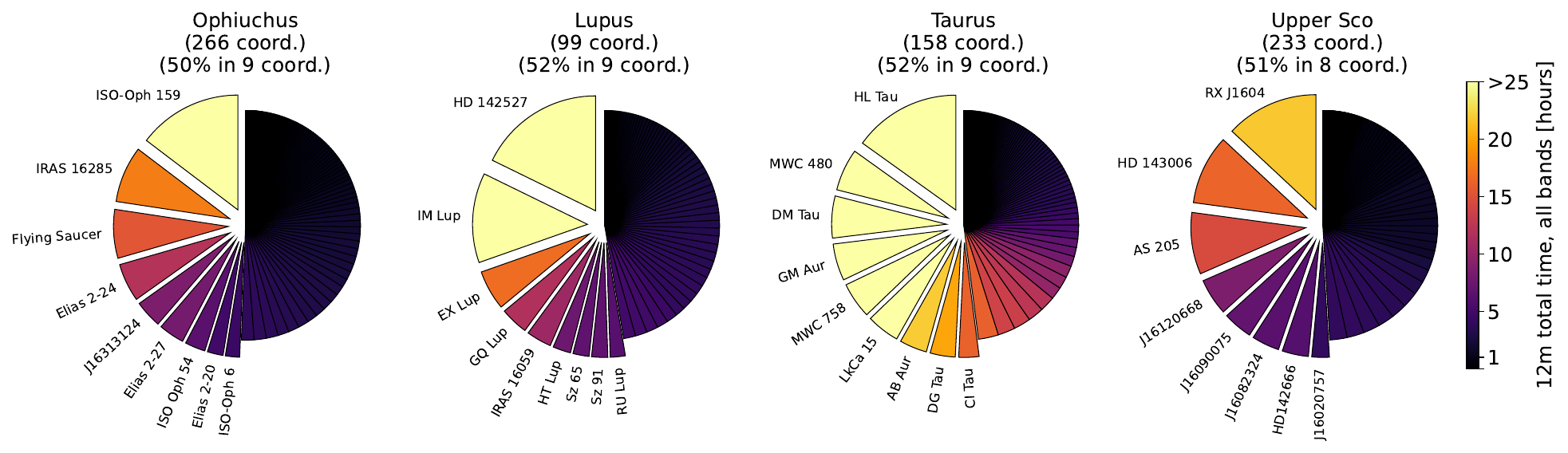}
    \caption{Distribution of time among sources in the most observed nearby star-forming regions. Targets that contribute to 50\% of the observed time of each region are annotated. The colorbar is saturated at 25\,hrs, representing sources with more accumulated time than the approximated threshold of an ALMA Large Program (1\,LPL). \label{fig:pie-time-per-source}}
    \vspace{4mm}
\end{figure*}

By matching the list of independent coordinates with those in the extended PPVII table, we can compare the distribution of exposure time across different star-forming regions, as presented in Fig.~\ref{fig:time-per-region}. Taurus and Lupus consistently accumulate the highest amount of time in nearly every ALMA Band. If we break down the time per region among their individual coordinates, we see that at least $50\%$ of all the time spent by ALMA in Taurus, Ophiuchus, Lupus, and Upper Sco, is distributed among $\leq9$ coordinates in each, as presented in Fig.~\ref{fig:pie-time-per-source}. Even though Taurus has the highest accumulated time in Bands 3, 6 and 7, most of this time can be attributed to a handful of sources, with 6 disks in Taurus accumulating $>1\,$LPL of data each. 

The region with most unequal distribution of time is Ophiuchus, where 3\% of the coordinates accumulate over $50\%$ of the time. Among these four regions, Upper Sco is the only one without a coordinate reaching $1\,$LPL. The most observed disk in Upper Sco is 2MASS J16042165-2130284 (also known as J1604 for short\footnote{In addition to the 2MASS name, another commonly used SIMBAD name for J1604 is ``RX J1604.3-2130A''. We warn the reader that if written without the final ``A'', the system could return incorrect sky coordinates, potentially leading to off-center pointings during observations.}), which is the 27th most observed system in the ``Disks and planet formation'' category (21.8 hrs, see Fig.~\ref{fig:hist-timesource}). This system is commonly used as a benchmark for disk shadows, which are detectable both in scattered light and at millimeter wavelengths with ALMA \citep{mayama2018, pinilla2015b, pinilla2018b, stadler2023}.

When considering the median accumulated time per source, Lupus and Taurus are leading with 22.4\,min and 9.5\,min per source in Band 6, respectively, ahead of the 1.5\,min per source in Ophiuchus. Upper Sco has been primarily observed in Band 7, where the median accumulated time per source is 2.4\,min, significantly below the 18.8\,min and 4.8\,min in the same band for Lupus and Taurus, respectively.

\begin{figure*}
    \centering
    \includegraphics[width=0.88\textwidth]{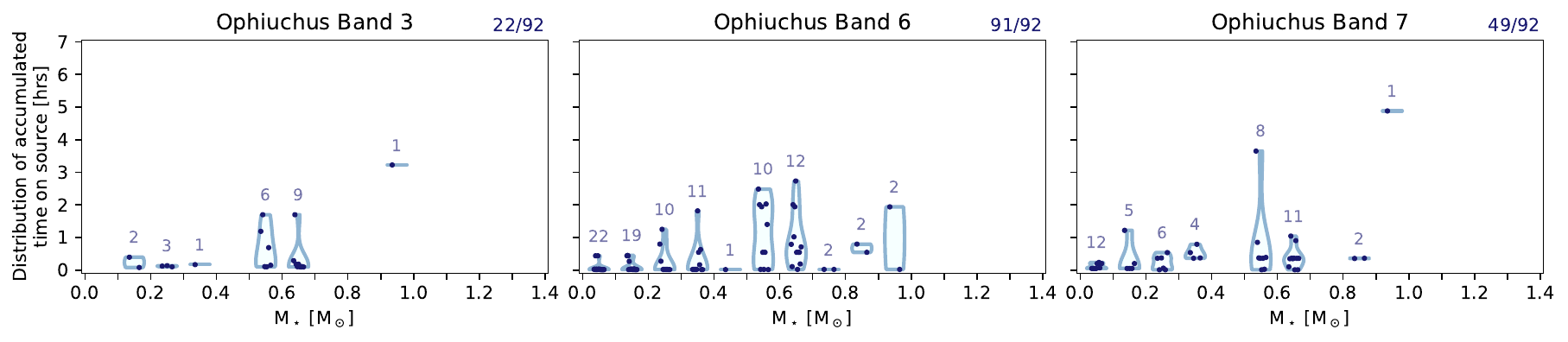}
    \includegraphics[width=0.88\textwidth]{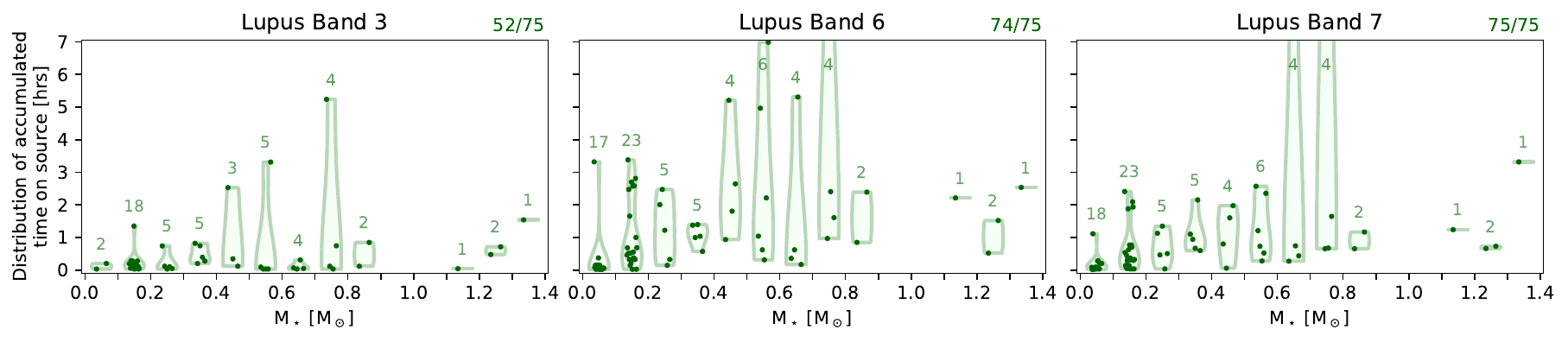}
    \includegraphics[width=0.88\textwidth]{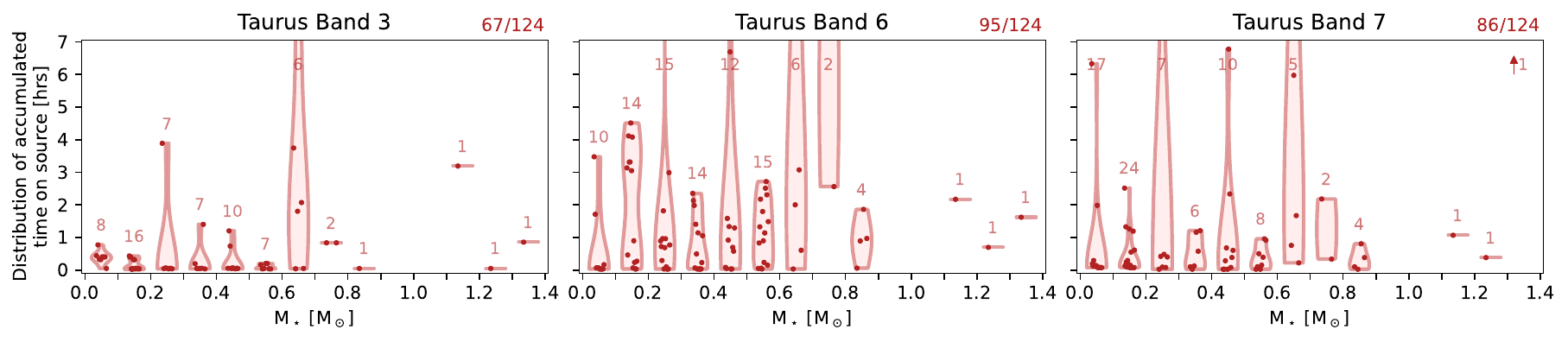}
    \includegraphics[width=0.88\textwidth]{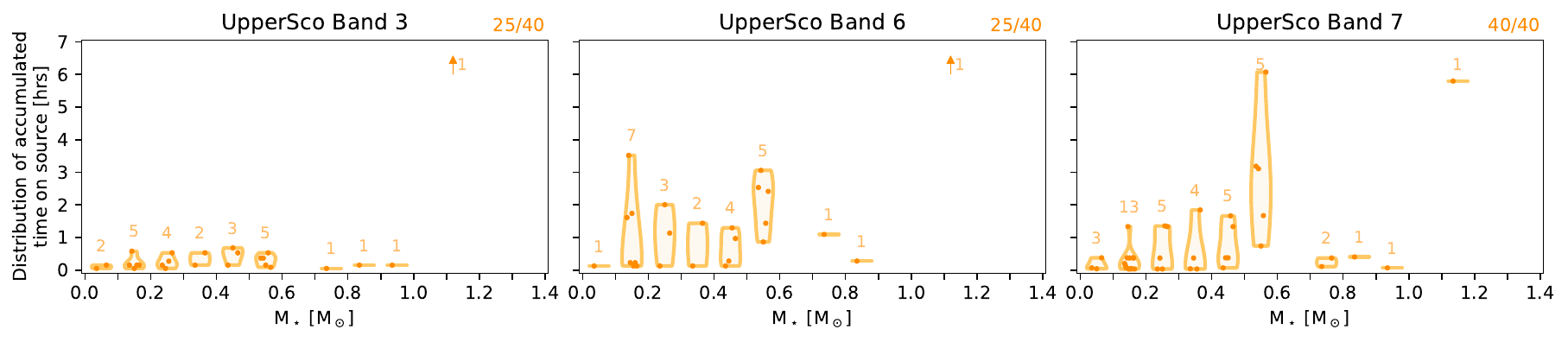}
    \includegraphics[width=0.88\textwidth]{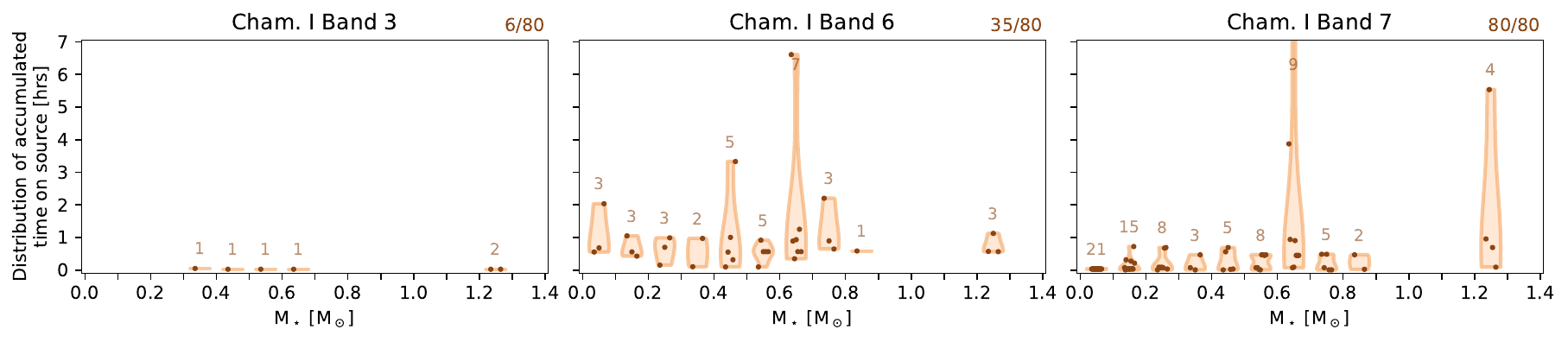}
    \caption{Distribution of accumulated exposure time per stellar mass bin. Only sources with known stellar masses are included in this figure. On the top right of each panel we show the fraction of sources with known stellar masses that have been observed in each band. The numbers above the distributions indicate the number of sources within that bin. The y-axis is limited to 7hrs, but a few disks go above this limit in Lupus, Taurus, Upper Sco and Chameleon I. The regions of Corona Australis and Chameleon II are in Fig.~\ref{fig:app:regions-time-per-band}. }
    \label{fig:regions-time-per-band}
\end{figure*}

\subsection{By stellar mass}

In Fig.~\ref{fig:regions-time-per-band}, we present the distribution of accumulated time in the three most observed ALMA Bands (3, 6, 7), separated by  region and stellar mass from the extended PPVII table. We note that only a fraction of sources are associated to a stellar mass in this table, as such constraints remain limited in the literature. 
Some of the most observed objects from Fig.~\ref{fig:hist-timesource} and \ref{fig:pie-time-per-source} contribute to extending the distributions above 7\,hrs on source for a few mass bins, particularly noticeable in Lupus and Taurus. 

We note a few patterns in the data: The stars in the range $M_\star > 0.2\,M_\odot$ are more commonly associated to longer accumulated time. For example, in Ophiuchus, most targets preferentially accumulate 1\,hr on source for stellar masses between $0.5\sim0.7\,M_\odot$, although a fraction of stars within that range are also consistent with only having snapshot observations, contributing to the unequal distribution of time in Ophiuchus mentioned in Sect.~\ref{sec:sec:by-sfr}. 

In Band 6, Lupus and Taurus show a bimodal distribution of accumulated time in the stellar mass bin from $0.1$ to $0.2\,M_\odot$, corresponding to very low mass stars (VLMS, see \citealt{pinilla2022b}). These coordinates are associated to the ALMA Large Program 2025.1.00324.L (DMOST), which is observing 7 VLMS disks in Lupus and 7 in Taurus, coinciding with the number of dots with $>1\,$hr of accumulated Band 6 time. Across all regions shown in Fig.~\ref{fig:regions-time-per-band}, 203 coordinates are consistent with being either a brown dwarf (BD) or a VLMS. Thus, the 14 DMOST sources with over 1\,hr in Band 6 represent less than 10\% of the population of very low mass objects (VLMOs, $M_\star<0.2\,M_\odot$). Band 3 is the frequency range where VLMOs are most under-observed.

\begin{figure*}[!t]
    \centering
    \includegraphics[width=0.99\textwidth]{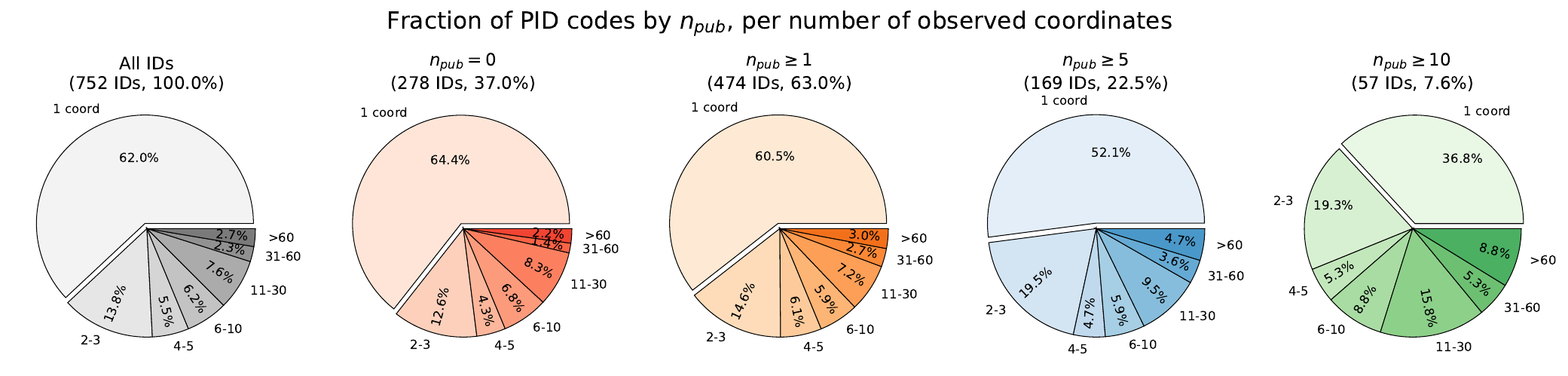}
    \caption{Distribution of Project ID (PID) codes by observed number of coordinates, as a function of associated number of publications (from zero to ten or more).}
    \label{fig:pie-numcoord-id}
    \vspace{4mm}
\end{figure*}

As discussed in Sect.~\ref{sec:sec:accumulated-exposure-time}, most of the coordinates have only been observed as part of snapshot programs, ranging from 30\,s to 360\,s of exposure time. Thus, we can use the accumulated time to roughly distinguish among targets observed only in snapshots, and those revisited in dedicated follow-up programs. We set the threshold at 12\,min per band, corresponding to the minimum exposure time dedicated by the ALMA Large Program 2022.1.00875.L (DECO), which is observing 80 disks across four nearby star-forming regions (Ophiuchus, Lupus, Taurus, Chameleon I). These results are presented in Fig.~\ref{fig:app:regions-completeness-per-band}. For lower stellar masses, the fraction of sources with dedicated observations decreases, with most VLMOs being consistent with only having been observed by snapshots. We note that this 12\,min threshold could originate from programs with different science goals, such as high angular resolution continuum or moderate angular resolution astrochemistry.

\subsection{By ALMA project code}\label{sec:sec:results-byPID}

We find 752 ALMA project codes in the scientific category of ``Disks and planet formation'', representing $13.7\%$ of the total number of project codes across all categories. In exposure time, they accumulate 2768.95\,hrs on source. The total time is divided into 2179.53\,hrs in Regular Programs, 66.49\,hrs in DDTs, 20.32\,hrs in Science Verification\footnote{We note that some Scientific Verification observations are not directly listed in the archive, such as the Band 4 observation of HL\,Tau.}, and 502.62\,hrs in Large Programs. These 752 project codes are dominated by single coordinate programs\footnote{The information of number of independent coordinates requested in the proposal of a project code is not publicly available. Incompletely observed programs might be registered in the archive with fewer coordinates than initially requested.} (i.e., a single target) with $62\%$ of the total, while $75.8\%$ of project codes observe three or fewer coordinates, as shown in the left panel of Fig.~\ref{fig:pie-numcoord-id}. Programs observing 10 or more coordinates account for only 12.6\% of the total number of project codes. Thus, the archive is dominated by observations taken as part of single target programs, contributing to an increased heterogeneity in sensitivity, frequency setups, and angular resolution. 

Of the 752 project codes, 278 (37\%) have no associated publication, while 474 (63\%) have at least one publication. From Fig.~\ref{fig:pie-numcoord-id}, we find no major difference between the distribution of coordinates in published or unpublished projects, as both trace the original distribution of project codes. However, programs with multiple coordinates contribute to a larger fraction of project codes that have received 5 or more publications. For programs with over ten publications, the distribution shifts toward being dominated by project codes with multiple coordinates. This suggests that the reusability value of a project code is higher for programs targeting more than one source. 

Earlier ALMA cycles have reached higher publication fractions, which is consistent with the findings of \citet{stoehr2026} when analyzing all the scientific categories of ALMA. We show these fractions in Fig.~\ref{fig:app:pub-fraction}, which show that $>80\%$ of project codes in a given cycle reach at least one publication, but it takes approximately 5–6 years to approach this level of completeness for a given ALMA Cycle. For reference, ALMA Cycle 6, which concluded observations in September 2019, has nearly reached the $80\%$ publication fraction and has reached about $50\%$ of project codes being used in 2 or more publications. This increasing fraction of project codes reaching multiple publications as a function of time is another supporting evidence for the long term value of the ALMA archive in the ``Disk and planet formation'' category. 

For ALMA Cycle 8, which began observations in October 2021 and concluded at the end of September 2022, project codes have not yet reached $50\%$ of publication rate. This delay is consistent with the median time of $4.2\,$yrs to reach one publication found by \citet{stoehr2026} when analyzing all science categories. 

Within the ``Disks and planet formation'' category, the project code with the highest number of publications is 2016.1.00484.L, corresponding to the DSHARP Large Program \citep{andrews2018b, huang2018b} awarded in Cycle 4, the first cycle with Large Programs. Since its publication and data release at the end of 2018\footnote{\url{https://almascience.eso.org/almadata/lp/DSHARP/}}, this project code has accumulated 62 publications, with only 10 of those corresponding to the first wave of articles from the collaboration responsible of the proposal. The top 20 project codes with the highest number of publications are presented in Tab.~\ref{tab:top20-numpub}, where we find four Large Programs (DSHARP; MAPS \citealt{oberg2021b}; eDisk \citealt{ohashi2023a}; and exoALMA \citealt{teague2025a}), one DDT program (early follow-up of PDS\,70, \citealt{keppler2019}), and one Science Verification program (high angular resolution observations of HL\,Tau, \citealt{almapartnership2015}). This top 20 is completed by 14 Regular programs, most of which are early surveys of disks, or first high sensitivity studies of typical benchmark sources (such as those from Fig.~\ref{fig:hist-timesource}). Almost all the Regular programs in this top are from ALMA Cycle 0-4, with the exception of 2019.1.01813.S, which was observed in ALMA Cycle 7. 

Similarly to Sect.~\ref{sec:sec:longest}, we estimate the publications per hour ratio of each project by dividing the number of publications to the total exposure time on source\footnote{Only for observations taken with the 12\,m array.}. We note that a single article may contribute to the publication count to several project codes. We find higher values of publication per hour ratio for earlier cycles, following the trend of Fig.~\ref{fig:app:pub-fraction}. Between Cycle 0 and 3 (before Large Programs), the median pub/hr per project code is $2.57_{-1.94}^{+5.26}$, where the reported range corresponds to the 16th and 84th percentile of the distribution. Between Cycle 4 and 9, the pub/hr for Regular programs is $0.76_{-0.76}^{+2.21}$, and for large programs is $0.72_{-0.45}^{+0.68}$ (including DSHARP, MAPS, eDisk, AGE-PRO, exoALMA). 
We note that the metric of pub/hr ratios is biased towards short exposure times, a topic we further discuss in Sect.~\ref{sec:sec:discussion-pubhr}.

\subsection{By spectral line}\label{sec:sec:line-emission}

\begin{figure*}
    \centering
    \includegraphics[width=1\textwidth]{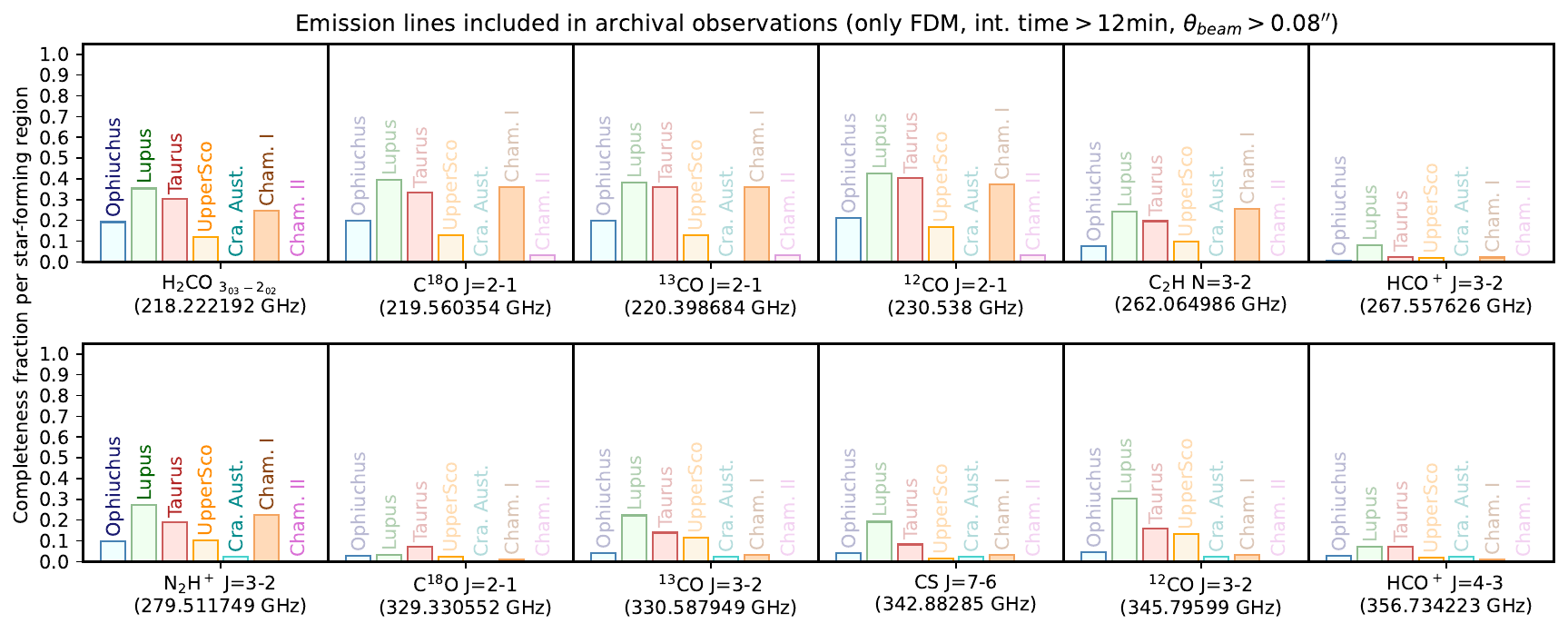}
    \caption{Completeness of emission lines observations per line, per  region. We include observations of at least 12\,min of exposure time, with a synthesized beam larger than $0.08''$, thus representing moderate or low angular resolutions. A fraction of 1.0 means that all the coordinates in a region have an observation including such emission line, where the total is defined from the number of coordinates from the extended PPVII table. }\label{fig:hist-lineobs}
    \vspace{0.2cm}
\end{figure*}

For each  region from the extended PPVII table, we evaluate the availability of moderate sensitivity observations of some commonly studied molecular lines. For each coordinate, we evaluate if the rest frequency of a molecular line is included in the frequency coverage of its observations. We exclude observations taken in Time Division Mode \citep[TDM,][]{escoffier2007}, as the broader frequency spacing would dilute line emission. We define moderate sensitivity as having more than 12\,min on source (the minimum time of the DECO molecular line survey, \href{https://almascience.eso.org/aq/?projectCode=2022.1.00875.L}{2022.1.00875.L}), and we filter by observations taken with low or moderate angular resolution, since high angular resolution produces lower brightness temperature sensitivity per beam\footnote{See \url{https://science.nrao.edu/facilities/vla/proposing/TBconv}}. These results are shown in Fig.~\ref{fig:hist-lineobs}, where we include molecular lines typically targeted in Band 6 (upper panel), and Band 7 (lower panel).

\begin{figure*}
    \centering
    \includegraphics[width=0.88\textwidth]{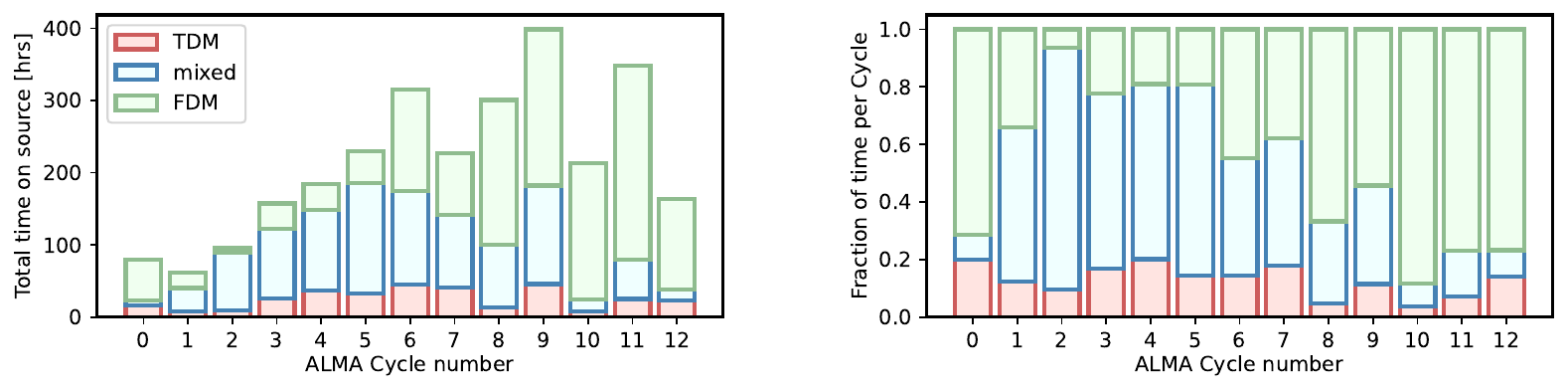}
    \caption{Fraction of time observed in full TDM (all spectral windows in TDM), mixed (spectral windows in TDM or FDM mode in the same observation), and FDM (all spectral windows in FDM), only for observations in the ``Disks and planet formation'' scientific category. Left panel shows total time per cycle, right panel is normalized by the total time. }\label{fig:hist-tdm-fdm}
    \vspace{0.2cm}
\end{figure*}

Among the emission lines included in Fig.~\ref{fig:hist-lineobs}, the Lupus region achieves the highest completion fraction for moderate sensitivity observations in Band 6. We note that the lines included in Fig.~\ref{fig:hist-lineobs} require several spectral setups to be observed. For example, the CO isotopologues in Band 6 are not compatible with observations targeting C$_2$H at 262.064986\,GHz, and neither with the N$_2$H$^+$ emission at 279.511749\,GHz. The higher completion fraction for the Band 6 lines and N$_2$H$^+$ can be attributed to the combined observations of the ALMA Large Programs 2021.1.00128.L (AGE-PRO) and 2022.1.00875.L (DECO), which targeted Ophiuchus, Lupus, Taurus, Upper Sco, and Chameleon I. 

The PI of each program has the choice of observing in TDM or in Frequency Division Mode (FDM) for each spectral window. While the TDM provides a slightly larger continuum bandwidth, the reduced frequency resolution is a limitation to detecting emission lines. In Fig.~\ref{fig:hist-tdm-fdm}, we show the distribution of time for projects in the ``Disks and planet formation category'' that observed either in full TDM, a mix of TDM and FDM, or full in FDM. Over time, a larger fraction of the time per cycle has been taken in full FDM, with steep increases in Cycle 6, 8, and 10. Observations using TDM for at least one spectral window reached an all time low in Cycle 10, representing only $12\%$ of the total time on source, and it later increased to $23\%$ in Cycle 11.

\begin{figure*}[t!]
    \centering
    \includegraphics[width=0.88\textwidth]{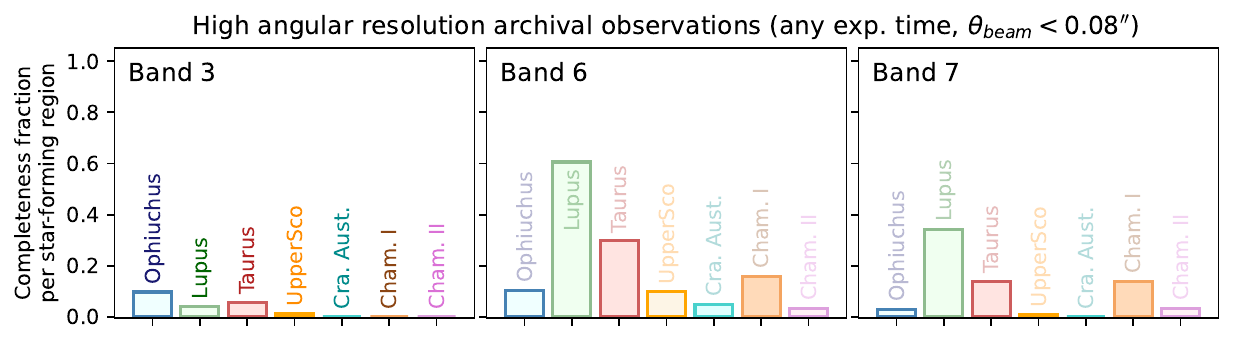}
    \includegraphics[width=0.88\textwidth]{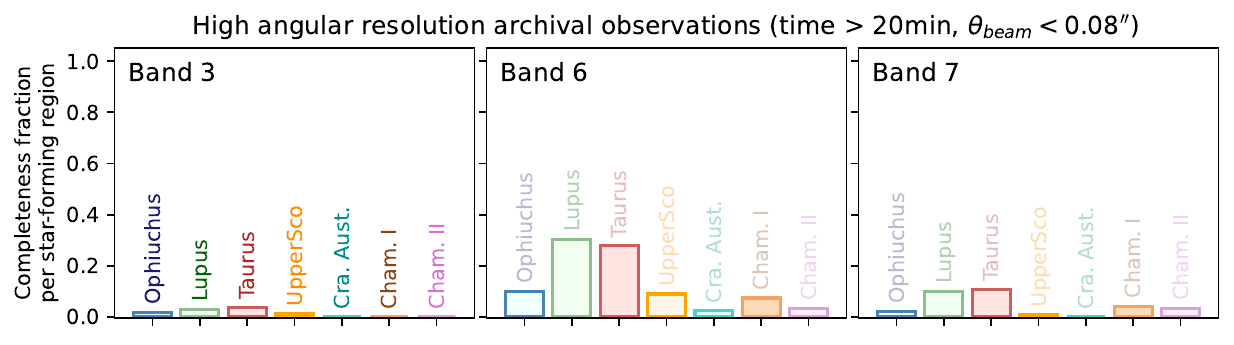}
    \includegraphics[width=0.88\textwidth]{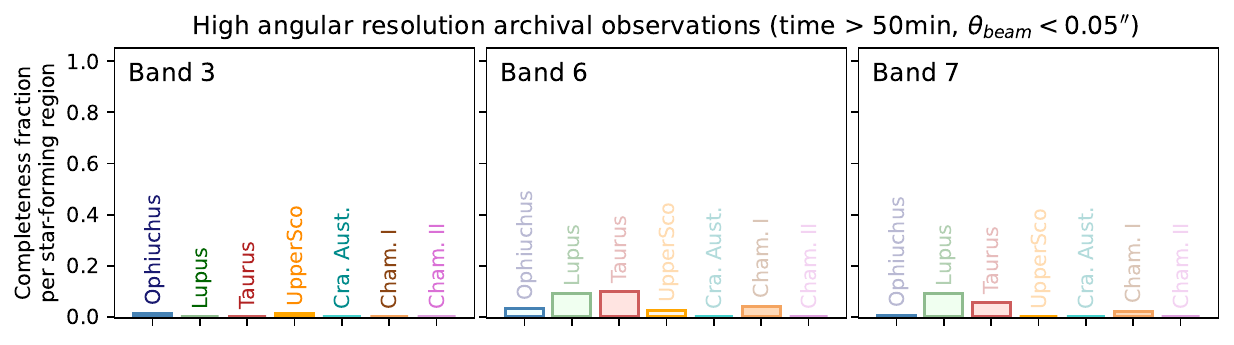}
    \caption{Completeness of high angular resolution observations (beam smaller than $0.08''$) in dust continuum emission, in the most popular ALMA Bands. A fraction of 1.0 means that all the coordinates in a region have high angular resolution observations. In the upper panel, all observations are considered, regardless of time on source. In the middle panel, a cut is applied at 20\,min to filter for moderate sensitivity observations. In the lower panel, we filter for observations roughly comparable to the DSHARP setup (see \citealt{andrews2018b}).}
    \label{fig:hist-highres}
    \vspace{0.2cm}
\end{figure*}

\subsection{By angular resolution}

While ALMA tested its capability for very high angular resolution during Cycle 2 \citep{almapartnership2015}, the longest baseline configurations only became standard\footnote{See \href{https://almascience.eso.org/observing/observing-configuration-schedule/prior-cycle-observing-and-configuration-schedule}{12m antenna configuration schedule}} during Cycle 4, when the equivalent to C43-9 and C43-10 were used for Regular and Large programs. These configurations (or their equivalent in baseline length) have been offered in Cycles 4, 5, 6, 7, 9, and 11. We invert the angular resolution condition utilized in Sect.~\ref{sec:sec:line-emission}, and we search for all observations with higher angular resolution than $0.08''$, and we present those findings as completion fractions in Fig.~\ref{fig:hist-highres}. 

Lupus is the region with the highest completion of high angular resolution, when considering observations with any exposure time. Lupus high angular resolution observations were recently analyzed as a sample by \citet{guerra-alvarado2025}, where substructures were found ubiquitous among spatially resolved disks. The second region with highest completion fraction is Taurus, recently studied by \citet{yamaguchi2024}. For Ophiuchus, high angular resolution observations are included in \citet{cieza2021} and \citet{shoshi2025}. For Upper Sco, a recent analysis of substructures was done by \citet{pinilla2025a} based on the moderate resolution survey of \citet{carpenter2025a}. 

While snapshots at high angular resolution are useful for spatially resolving the disk's emission, they are also limited to detect high contrast features. If we filter by $>20$\,min on source, for enhanced brightness temperature sensitivity, we only see a decrease in completion fractions for Lupus in Band 6 and Ophiuchus in Band 3, while the other regions remain mostly constant in their fractions. If we filter by high sensitivity and very high angular resolution, comparable to DSHARP (35\,min to 70\,min per source, and beam size $\leq 0.05''$, \citealt{andrews2018b}), we see a steep decline in the completion fraction of all regions, as presented in the bottom panel Fig. \ref{fig:hist-highres}. 

Only two ALMA Large Programs have been dedicated to very high angular resolution observations (20 systems in DSHARP, \citealt{andrews2018b}; 19 systems in eDisk, \citealt{ohashi2023a}). Therefore, the majority of high angular resolution observations have been obtained as part of Regular programs, which contributes to increasing the heterogeneity of available angular resolutions, sensitivity, and representative wavelengths. In Sect.~\ref{sec:app:galleries}, we present a non-exhaustive summary of notable dust continuum substructures identified in Class II disks, such as high-contrast wide asymmetries in Fig.~\ref{fig:disks_large_asym}, localized asymmetries in Fig.~\ref{fig:disks_localized_asym}, spirals in Fig.~\ref{fig:disks_spirals}, and spatially resolved structured disks in Fig.~\ref{fig:disks_rings}. Further detail is given in Sect.~\ref{sec:app:galleries}. 
For a dedicated sample of central cavities, we refer the reader to \citet{vioque2026}. For young Class 0/I objects, recent sample studies were published by \citet{maureira2026} and \citet{ohashi2023a}. For debris disks, we refer to \citet{han2026, marino2026} and \citet{matra2025}.

\section{Discussion}\label{sec:discussion}

\subsection{The distribution of project codes among coordinates}

The three main mechanisms to obtain ALMA data are Regular programs, Large programs, and Director's Discretionary Time (DDT). After almost 15 years of operations, the archive of ``Disks and planet formation'' is now composed of 752 project codes, of which 10 are ALMA Large Programs (DSHARP, MAPS, eDisk, AGE-PRO, exoALMA, ARKS, DECO, CHEER, DiskStrat, DMOST), 721 are Regular Programs, 18 are DDT, and 3 are Science Verification (SV). By comparing the integrated exposure time per project code, ALMA Large Programs only account for $18.5\%$ of the total exposure time. Thus, the majority of the archival data is a collection of observations from Regular programs, which will dominate the distribution of projects among coordinates. 

While the distribution of observed projects is dependent on the distribution of proposed projects, there are different processes that will modify how proposal pressure translates into observations. Proposals are first reviewed, then accepted based on their ranking, and finally scheduled into the observing queue, and each of these steps could introduce modifications or biases. 

Since Cycle 8, Regular programs go through the Distributed Peer Review \citep[DPR,][]{carpenter2022, carpenter2025b, donovan2022}, where every proposal is ranked by 10 anonymous reviewers among the community. This mechanism is split into two stages: Stage 1, when reviewers submit their rankings, and Stage 2, when reviewers have the chance of modifying their rankings. As this is a distributed mechanism, there is no unique directive about what should be the observing strategy. Thus, the most highly ranked proposals of each cycle reflect a combination of the individual assessments of all reviewers. 

After being reviewed, proposals are accepted based on their scheduling feasibility and distribution of time among the ALMA partner regions. Therefore, proposals requesting time in oversubscribed regions of the sky may have a lower acceptance rate due to the limited time available per Right Ascension (RA). While the proposal pressure per RA in each cycle is not publicly available, we can estimate the RA ranges with the highest concentration of archival observations, as presented in Fig.~\ref{fig:app:hist-time-per-ra}. For the ``Disks and planet formation'' category, observations are mostly obtained in the RA of the nearby star-forming regions, particularly at the RA of Taurus (RA$\approx$4\,hrs 30\,min), and for Lupus, Ophiuchus, and Upper Sco (RA$\approx$16\,hrs).

\begin{figure}
    \centering
    \includegraphics[width=0.46\textwidth]{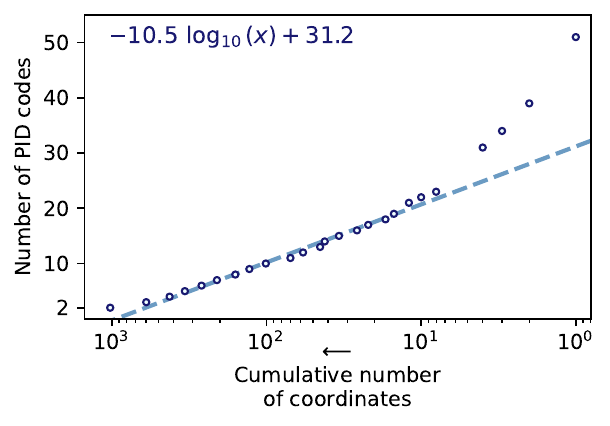}
    \caption{Distribution of number of independent project codes accumulated by source. About $\sim1000$ coordinates have been observed by at least 2 project codes, $\sim100$ coordinates have been observed by at least 10 project codes, and only one coordinate has more than 50 project codes. Here we only present coordinates with at least 2 project codes. The cumulative number of coordinates with at least one project code is 3933. }
    \label{fig:numids-coord-log}
    \vspace{0.2cm}
\end{figure}

Even if a proposal is accepted, its observability will depend on the observing queue building process\footnote{Described in the ALMA Technical Handbook through the dynamic scheduling software \citep{alma_technicalhandbook_cycle13}.}, which will directly depend on weather conditions, elevation of a coordinate, and completion of a program. Due to the challenges of observing in the high frequency Bands, this last step is going to lead to different oversubscription rates as a function of Band (see also ESO 2024 Annual Report, \citealt{eso_ar2024}, page 49). 

It is yet to be quantified how the original distribution of proposed programs is modified by each of these steps (review, acceptance, scheduling), but a comparison between the distribution of programs in samples, coordinates, and frequency setups, could contribute to identifying biases or challenges for obtaining different types of observations. In the specific case of the ``Disks and planet formation'' category, the emergent distribution after all these steps is dominated by single coordinate programs ($62\%$ of all project codes, see Fig.~\ref{fig:pie-numcoord-id}). 

We can better understand the emergent observing distribution by calculating the cumulative distribution of project codes per coordinate. In other words, how many different project codes have observed the same coordinate. The observed distribution is well described with a single logarithmic function, except for the four coordinates with the highest number of project codes, as presented in Fig.~\ref{fig:numids-coord-log}. If we take the number of project codes as an indication of the interest of the community on a particular coordinate, then we find that the community is considerably more interested in a handful of coordinates, or the inverse of the argument, it is less interested in a random coordinate taken from the sample following a logarithmic distribution. There are four coordinates above the trend, corresponding to TW\,Hya, HD\,163296, IRAS\,16293-2422, and IRAS\,04368+2557. These sources accumulate over 30 project codes each, and also coincide with some of the most observed coordinates in Fig.~\ref{fig:hist-timesource}.

\begin{figure}
    \centering
    \includegraphics[width=0.46\textwidth]{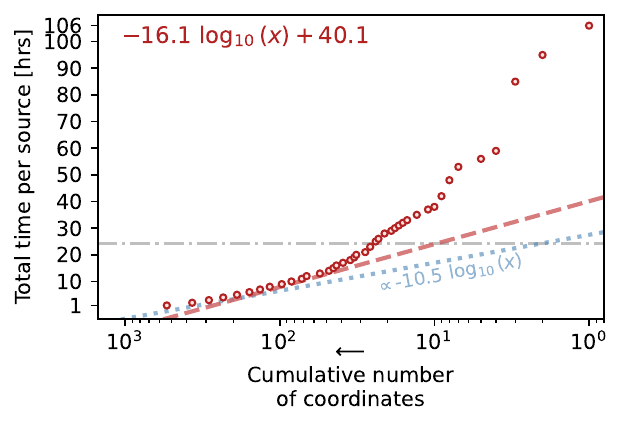}
    \caption{Distribution of time accumulated by source. For reference, the dashed dotted horizontal line shows the total time of project code 2016.1.00484.L, and the dotted blue line shows the slope of the distribution of project codes per coordinate. The top 50 sources (from right to left) were previously presented in Fig.~\ref{fig:hist-timesource}. This panel only shows coordinates with at least 1\,hr on source. The cumulative number of coordinates with more than zero hours of exposure time is 3933. }
    \label{fig:times-coord-log}
    \vspace{0.2cm}
\end{figure}

\begin{figure*}
    \centering
    \begin{minipage}{0.92\textwidth}
        \begin{minipage}{0.49\textwidth}
            \centering
            \includegraphics[width=\textwidth]{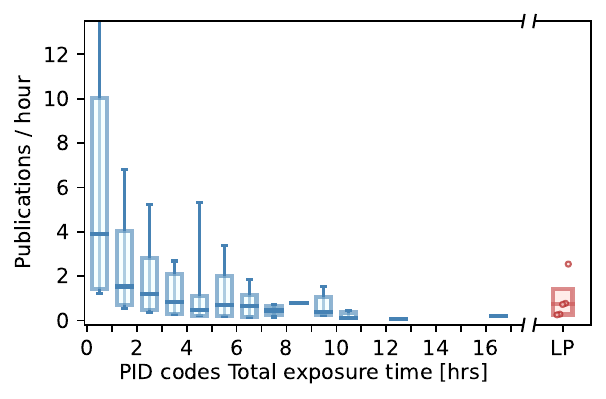}
        \end{minipage}
        \hfill
        \begin{minipage}{0.49\textwidth}
            \centering
            \includegraphics[width=\textwidth]{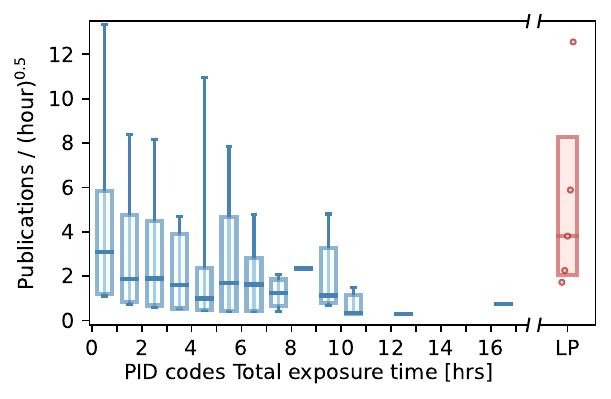}
        \end{minipage}
    \end{minipage}
    \caption{Distribution of publication per time ratio, as a function of the total exposure time of a project code. Left panel shows publications per hour, and right panel shows publications per square root of hour. The rectangles show the 68 percentile for each bin, and the errorbars show the 97.5 percentile. 
    We only included project codes between ALMA Cycle 4 and Cycle 9, and we exclude project codes with no publications. The five Large Programs between Cycle 4 and 9 are shown on the right side of each panel. }
    \label{fig:numpubhr-numpubsqrt}
    \vspace{0.2cm}
\end{figure*}

\subsection{Distribution of time among disks}\label{sec:sec:dist-time-sources}

We present the cumulative distribution of time among coordinates in Fig.~\ref{fig:times-coord-log}. We observe a large deviation from a single logarithmic distribution, where the top $\approx100$ most observed coordinates have a steeper distribution than the rest of the sample. If all project codes observed their coordinates for exactly the same amount of time, we would expect the distribution of time of Fig.~\ref{fig:times-coord-log} to follow the same trend as in Fig.~\ref{fig:numids-coord-log}. 

In addition to the preference for observing a group of coordinates with more project codes, there is also a preference for observing those coordinates with project codes of longer exposure time. This could be interpreted as using these sources as benchmarks, dedicating more time to them to achieve higher sensitivity, and to later extrapolate their results to the general disk population. In this context, TW\,Hya is the top benchmark for studies of planet forming disks, both in number of project codes and in exposure time. The proximity of TW\,Hya allows for studies that would be more challenging or expensive in other sources. In addition, TW\,Hya is conveniently located at a RA with lower stress than the most observed nearby star-forming regions (see Fig.~\ref{fig:app:hist-time-per-ra}). We highlight that even after a decade and a half of observations, the community continues to find new features and test novel methods on TW\,Hya \citep[e.g.,][]{ilee2022, das2024, teague2025b, yoshida2022}. A similar science case can be made for other objects in the top 50 ranking. However, it is important to identify science cases which are underrepresented among the most observed objects. For example, there is no VLMS in the top 50 ranking of accumulated time, while also remaining underobserved as a population (see Fig.~\ref{fig:app:regions-completeness-per-band}). On the group of objects accumulating more time than 1 LPL, we only find two circumbinary disks (HD\,142527, V4046\,Sgr) and the young object BHR\,71 is the only multiple disk system\footnote{Multiple disk systems refer to stellar systems where each component has its own independent disk, differently from a circumbinary disk.} among these top 24 (composed by IRS 1 and IRS 2, a very wide separation binary). Even though most stars are part of multiple stellar systems, the first Class II binary disks systems appear only at ranking 41 and 46, for T\,Tau and AS\,205, respectively. 

A similar benchmark strategy is observed in the nearby star-forming regions, with a small subset of systems ($\approx 9$) accumulating half of the region's time. These sources are typically bright single disk systems. In the multiple disks systems included in Fig.~\ref{fig:pie-time-per-source}, we find HT\,Lup, Sz\,65, and AS\,205, for which their missing orbital parameters are a challenge to studying their dynamical interactions \citep[see][]{weber2023, jorquera2024, phuong2026}. While bright and extended disks in single stellar systems provide important benchmarks for the study of planet formation, they might also bias theoretical interpretations. It is important to consider that the majority of the Class II disk population is composed of compact disks \citep[e.g.,][]{guerra-alvarado2025}, and most of the stars are of very low mass.

\subsection{Efficiency of publications per hour}\label{sec:sec:discussion-pubhr}

Recent reports from the ALMA Scientific Advisory Committee (ASAC) have mentioned interest in evaluating the scientific impact measured in publications or citations per unit of time for Regular and Large Programs \citep{tafalla2023, milam2024}, which was recently analyzed for all ALMA observations in \citet{stoehr2026} by comparing citations and publications per hour of ALMA. We estimated the number of publications per hour of exposure time (pub/hr) for the project codes of the ``Disks and planet formation'', and we present them in bins of exposure time in Fig.~\ref{fig:numpubhr-numpubsqrt}. 

We note a correlation between pub/hr and the total exposure time, resulting in bias of larger pub/hr ratios for shorter project codes (less than 1\,hr on source). As an example, an observation of 20\,min with one publication would be registered as having the same pub/hr as another with 6\,hrs on source with 18 publications. Thus, pub/hr tends to be elevated for short exposure times, and sets a challenging standard for project codes aiming for higher sensitivity, required for a variety of science goals. It is beyond the scope of this work to verify if the metric of citations per hour is also biased towards shorter observations, but we note that an analysis of the Hubble Space Telescope productivity by \citet{apai2010} also found larger values of citations per orbit for programs requesting fewer orbits. 

As an experiment, we checked the distribution of publications per square root of hours ($\text{pub}/\sqrt{\text{hr}}$), to take into consideration that the thermal sensitivity of an observation scales inversely as the square root of the exposure time ($\sigma_{\text{rms}}^{-1}\propto\sqrt{\text{time on source}}$). These results are presented for project codes between Cycle 4 and Cycle 9 in the right panel of Fig.~\ref{fig:numpubhr-numpubsqrt}. While this metric should be characterized in further detail, we note that the median $\text{pub}/\sqrt{\text{hr}}$ is flattened for project codes above 1\,hr on source, and their 68 percentile distributions become consistent among each other.

While Regular and Large Programs have a similar pub/hr ratio when measured over a similar time period (see Sect.~\ref{sec:sec:results-byPID}), the metric of $\text{pub}/\sqrt{\text{hr}}$ shows that Large Programs outperform the median of other project codes, enabling a larger number of research articles per investment of telescope time. This is similar to our finding in Sect.~\ref{sec:sec:longest}, where ALMA observations with the longest exposure times are utilized by a larger number of publications, when taken in the context of a Large Program or community efforts. Between Cycle 4 and Cycle 9, we find a $\text{pub}/\sqrt{\text{hr}}$ of $3.80_{-1.74}^{+4.48}$ for Large Programs, $2.96_{-2.32}^{+2.46}$ for DDTs, and $1.11_{-1.11}^{+2.82}$ for Regular Programs.

\section{Conclusions}\label{sec:conclusions}

ALMA has observed planet forming disks for almost 15 years. These observations have become a large and heterogeneous survey of millimeter emission for objects at different stages of planet formation. The combined observations of the ``Disks and planet formation'' category are an achievement of the community as a whole, and we summarized them as a function of time on source, region, emission lines, and angular resolution. Our results are limited to February 2026. We provide our main conclusions as follows: 
\begin{itemize}
    \item ALMA has observed 3933 independent coordinates as part of the ``Disks and planet formation'' category, distributed over 752 independent project codes. These projects integrate to 2779.53\,hrs on source. 
    \item The main ALMA Bands in exposure time and number of coordinates are Band 3, 6, and 7. The higher frequencies of Band 9 and 10, which are more challenging to schedule, have only been used for $\sim1.4\%$ of the coordinates in the sample. 
    \item We identify 24 sources with accumulated exposure times longer than the threshold for an ALMA Large Program. These sources provide benchmarks to test theoretical predictions and analysis methods. 
    Multiple disk systems are less represented than single disk systems in Fig.~\ref{fig:hist-timesource} and Fig.~\ref{fig:pie-time-per-source}. No very low mass object (VLMS or BD) is found among the top 50 most observed coordinates, or top 10 of each region. 
    \item The longest single observations, at a specific frequency setup, are similarly distributed among debris disks and planet forming disks. When selecting only those with one or more associated publications, we find that the project codes taken as part of community efforts (Large Programs or Science Verification) are utilized in a larger number of publications than those from Regular Programs. 
    \item Lupus and Taurus are the most observed star-forming regions in exposure time, followed by Ophiuchus and Upper Sco. In these four regions, over 50\% of the time on source can be attributed to 9 coordinates each. 
    \item The completeness of line emission observations at moderate resolution is different between star-forming regions. Two Large Programs have provided the largest contribution to increasing completeness fractions in the emission lines of Band 6, as well as N$_2$H$^+$\,J=3-2 in Band 7. 
    \item Most observations in recent cycles are taken with all spectral windows in FDM, which allows for the characterization of molecular line emission in spectral windows normally dedicated to continuum emission. 
    \item Lupus has the highest completeness fraction in observations at high angular resolution. By filtering for moderate sensitivity, Lupus and Taurus have similar completeness of $\approx30\%$, while other regions are yet to reach $10\%$. High sensitivity and high angular resolution observations are not common for any star-forming region. 
    \item For coordinates observed by at least two project codes, the distribution of project codes per coordinate is logarithmic. By comparing with the distribution of total exposure time, we see that the community has dedicated more project codes and longer exposure times to a group of coordinates ($\approx$100), consistent with an observing strategy focused on benchmark targets. 
    \item Most programs are dedicated to a single coordinate ($62\%$ of the total), contributing to the heterogeneity of the ALMA archive. Project codes observing at least a small sample are more likely to be reused in several publications. 
    \item When measuring the publication efficiency of project codes, the metric of pub/hr tends to overrepresent observations with $<1\,$hr on source, setting a challenging standard for longer observations. We find that the metric of publications per square root of time contributes to reducing this bias, and provides a better comparison for project codes with different exposure times. 
    \item The median $\text{pub}/\sqrt{\text{hr}}$ of Large Programs outperforms the median of other programs, when binned in steps of 1\,hr of exposure time. The public availability of high level data products may be one of the explanations for this trend. 
\end{itemize}

The archive hosts a collection of observations for samples and single targets that would be unfeasible to obtain through any single program. We encourage the community to identify scientific questions that are becoming possible through this wealth of archival data.

\bigskip

\section*{Acknowledgements}

We are thankful to the ALMA staff and everyone involved in enabling the scientific operations of ALMA. We thank all the teams who made their published data publicly available, included in the citations of Tab.~\ref{tab:references_disks_1} and \ref{tab:references_disks_2}. We also thank J. Carpenter, E. Macias, F. Stoehr, and A. Ziampras for the useful discussions. 
L.P. acknowledges support from ANID BASAL project FB210003 and ANID FONDECYT Regular \#1262272.

\clearpage

\appendix

\section{A. Most published ALMA project codes}

In Tab.~\ref{tab:top20-numpub}, we list the Project codes with the highest number of publications (to the beginning of February 2026). We note that the number of publications is constantly changing, and therefore the table should be revised regularly. 

\begin{table}[h!]
    \centering
    \caption{Top 20 ALMA project codes with the highest number of publications.}
    \begin{tabular}{l|c|c|c|c|c|c|l|c}
N  & pos. &  project     & Num. & Num.  & pub/hr & Num.   & Title            & Related \\
   & (*)  &  code        & pub. & hr    & ratio  & coord. & (shortened)      & publications \\
\hline\hline
1  & 1  & 2016.1.00484.L  & 62   & 24.38 &  2.54 &  20 & Small-Scale Substructures in Protoplanetary Disks     & 1, 2       \\
2  & 2  & 2013.1.00498.S  & 47   &  2.20 & 21.36 &  12 & Dust growth in protoplanetary disks: where in the ... & 3, 4       \\
3  & 3  & 2018.1.01055.L  & 45   & 58.50 &  0.77 &   5 & The Chemistry of Planet Formation                     & 5, 6       \\
4  & 4  & 2016.1.01164.S  & 37   &  4.10 &  9.03 &  32 & An unbiased survey of disk structures in Taurus       & 7, 8, 9    \\
5  & 5  & 2011.0.00015.SV & 31   & 16.29 &  1.90 &   1 & Science verification observation of HL Tau            & 10         \\
6  & 6  & 2015.1.00888.S  & 26   &  3.41 &  7.62 &   3 & Probing disk structure in a cavity of ..              & 11, 12, 13 \\
7  & 7  & 2013.1.00601.S  & 24   &  2.47 &  9.72 &   1 & ALMA measurements of disk turbulence                  & 14         \\
8  & 8  & 2016.1.00344.S  & 22   &  5.62 &  3.92 &   2 & Detecting the kinematical signature of accreting ...  & 15         \\
9  & 9  & 2013.1.00220.S  & 21   &  2.80 &  7.51 &  95 & Disk Demographics in Lupus                            & 16         \\
10 & 10 & 2015.1.00678.S  & 20   &  5.14 &  3.89 &   6 & Survey of CO Snow Lines in Solar Nebula Analogues     & 17, 18, 19 \\
11 & 10 & 2019.1.00261.L  & 20   & 27.65 &  0.72 &  16 & Early Planet Formation in Embedded Disks              & 20         \\
12 & 12 & 2015.1.00486.S  & 19   &  1.31 &  14.5 &   1 & On the Gas-to-Gust Mass ratio in T Tauri Disks        & 21         \\
13 & 13 & 2015.1.00686.S  & 18   &  2.20 &  8.19 &   1 & Characterizing Substructure in the TW Hya Disk        & 22         \\
14 & 13 & 2012.1.00158.S  & 18   &  2.05 &  8.78 &   3 & Quantifying gas inside dust cavities in ...           & 23, 24     \\
15 & 13 & 2021.1.01123.L  & 18   & 64.01 &  0.28 &  15 & exoALMA                                               & 25         \\
16 & 16 & 2011.0.00526.S  & 17   &  2.15 &  7.91 &  20 & A Survey of Circumstellar Disks around Low-mass ...   & 26, 27     \\
17 & 16 & 2016.1.00115.S  & 17   &  2.22 &  7.67 &   1 & Disentangle the polarization mechanims between  ...   & 28         \\
18 & 16 & 2013.1.00658.S  & 17   &  0.39 & 43.06 &   2 & Hunting for gaps in HEABE disks                       & 29, 30     \\
19 & 19 & 2017.A.00006.S  & 16   &  1.88 &  8.49 &   1 & First detection of a circumplanetary disk within ...  & 31         \\
20 & 19 & 2019.1.01813.S  & 16   &  9.92 &  1.61 & 882 & SODA: a flux-limited Survey of Orion's Disks ...      & 32         \\
\hline\hline
    \end{tabular}
      \begin{minipage}{16cm}
        \vspace{0.1cm}
        (*) Project codes with the same number of publications are ranked at the same position in the second column. \vspace{0.2cm} \\
        References: 
        (1)  \citet{andrews2018b}, 
        (2)  \citet{huang2018b}, 
        (3)  \citet{perez2016}, 
        (4)  \citet{pinilla2018a}, 
        (5)  \citet{oberg2021b}, 
        (6)  \citet{law2021}, 
        (7)  \citet{long2018b}, 
        (8)  \citet{long2019}, 
        (9)  \citet{manara2019}, 
        (10) \citet{almapartnership2015}, 
        (11) \citet{longzachary2018}, 
        (12) \citet{mayama2018}, 
        (13) \citet{keppler2019}, 
        (14) \citet{isella2016}, 
        (15) \citet{perez2019}, 
        (16) \citet{ansdell2016}, 
        (17) \citet{kastner2018}, 
        (18) \citet{qi2019}, 
        (19) \citet{vandermarel2019}, 
        (20) \citet{ohashi2023a}, 
        (21) \citet{fedele2018}, 
        (22) \citet{andrews2016}, 
        (23) \citet{pinilla2015a}, 
        (24) \citet{vandermarel2016}, 
        (25) \citet{teague2025a}, 
        (26) \citet{carpenter2014}, 
        (27) \citet{barenfeld2016}, 
        (28) \citet{kataoka2017}, 
        (29) \citet{vanderplas2017a}, 
        (30) \citet{vanderplas2017b}, 
        (31) \citet{keppler2019}, 
        (32) \citet{vanterwisga2022}.
      \end{minipage}
    \label{tab:top20-numpub}
\end{table}

\clearpage

\section{B. Additional Figures.}\label{sec:app:additional_figs}

\begin{figure*}[h]
    \centering
    \begin{minipage}{0.92\textwidth}
        \begin{minipage}{0.49\textwidth}
            \centering
            \includegraphics[width=\textwidth]{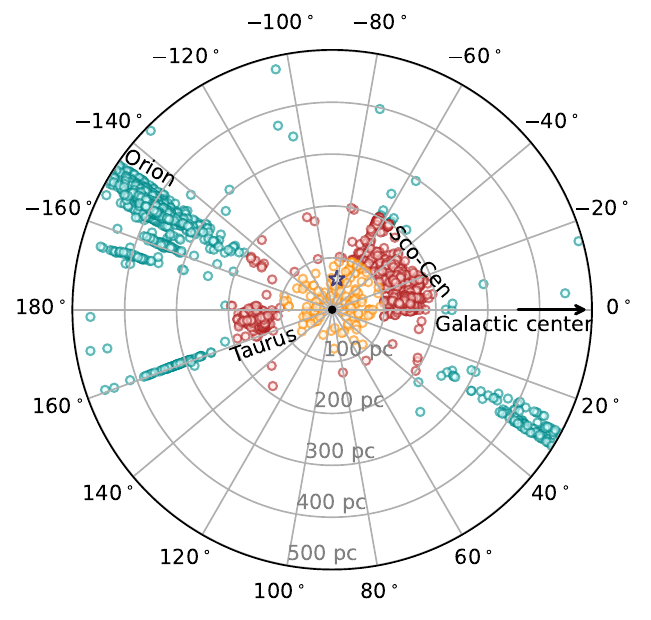}
            \caption{Distribution of observed coordinates in the galactic plane, with the galactic center to the right. The objects are color coded by the same distances as in the panels of Fig.~\ref{fig:sky-distance-all}. A star marker shows the location of TW\,Hya. The angles are the same as in Fig.~\ref{fig:sky-comb-coordinates}. \label{fig:app:galactic_plane}}
        \end{minipage}
        \hfill 
        \begin{minipage}{0.49\textwidth}
            \centering
            \includegraphics[width=\textwidth]{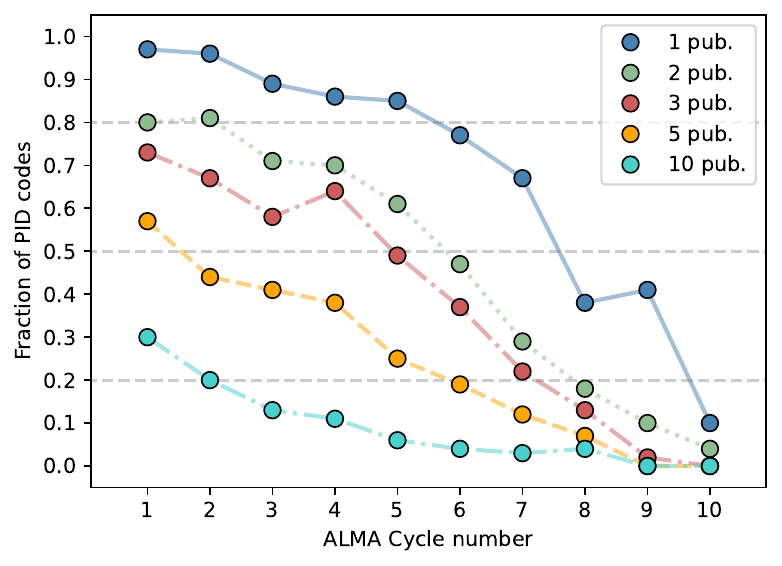}
            \caption{Publication fraction for projects in the ``Disks and planet formation'' category, as a function of ALMA Cycle. The dashed lines in the background represent the 20\%, 50\% and 80\% relative to the total number of projects for that cycle. \label{fig:app:pub-fraction}}
        \end{minipage}
    \end{minipage}
    \vspace{0.2cm}
\end{figure*}

\begin{figure*}[h]
    \centering
    \includegraphics[width=0.99\textwidth]{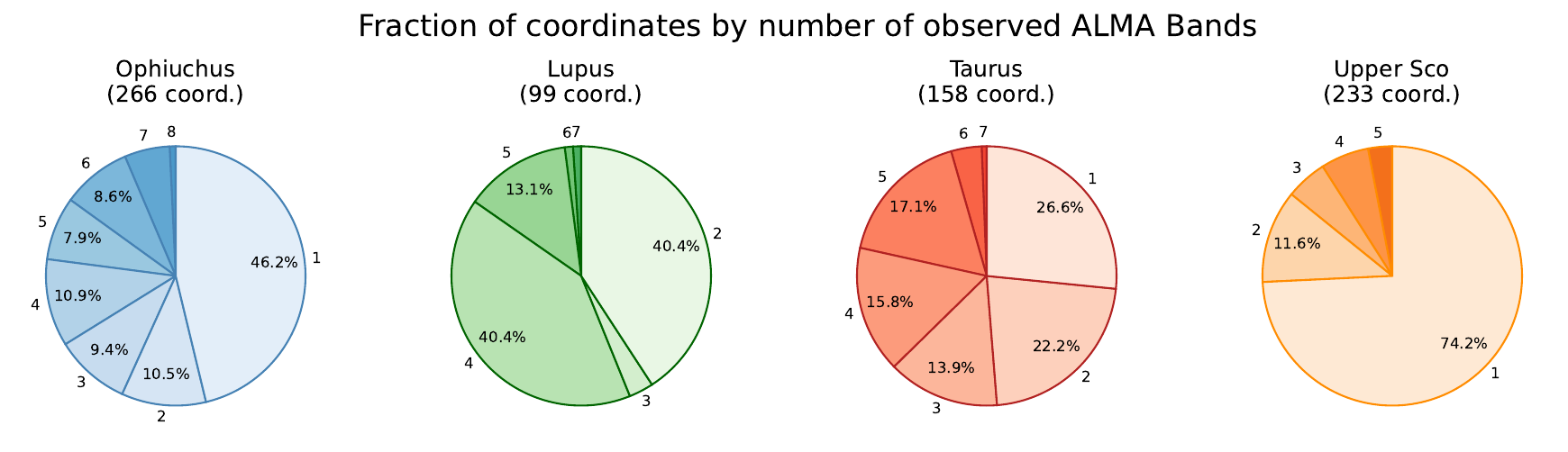}
    \caption{Number of ALMA Bands available for the coordinates of nearby star-forming regions. For example, 46.2\% of coordinates in Ophiuchus have observations in only one Band, while 10.5\% have observations in two Bands. }
    \label{fig:app:pie_bands-source}
\end{figure*}

\begin{figure*}[h]
    \centering
    \includegraphics[width=0.98\textwidth]{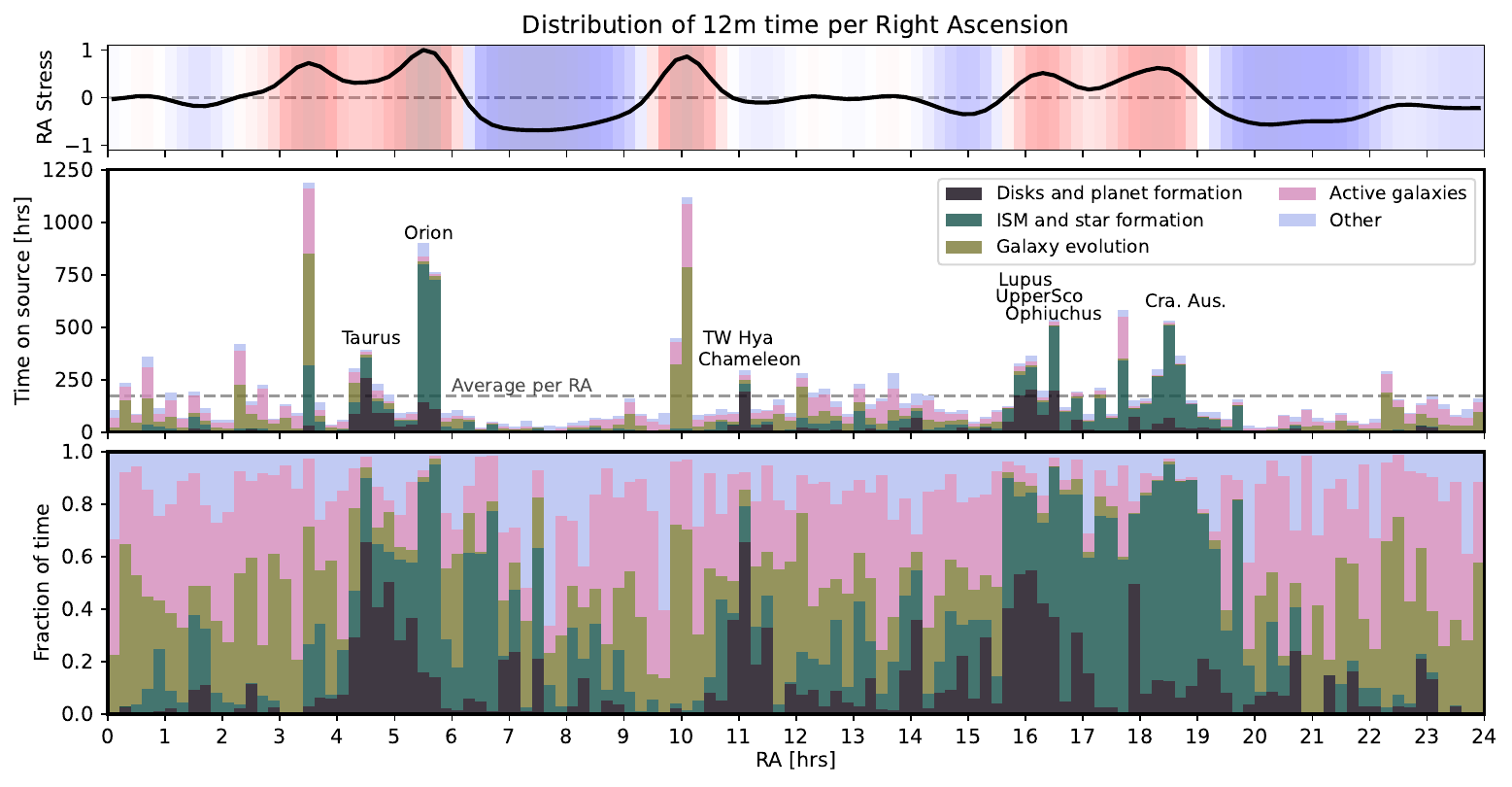}
    \caption{Distribution of time per Right Ascension in bins of 3\,deg (12\,min of RA), separated by scientific category. The top panel shows the RA stress, which peaks at RA=10\,hrs, and has two minimum between 7-9\,hrs and 19-22\,hrs. The lower panel shows the fraction of time per RA spent on each scientific category. }
    \label{fig:app:hist-time-per-ra}
\end{figure*}

\begin{figure*}[h]
    \centering
    \includegraphics[width=0.9\textwidth]{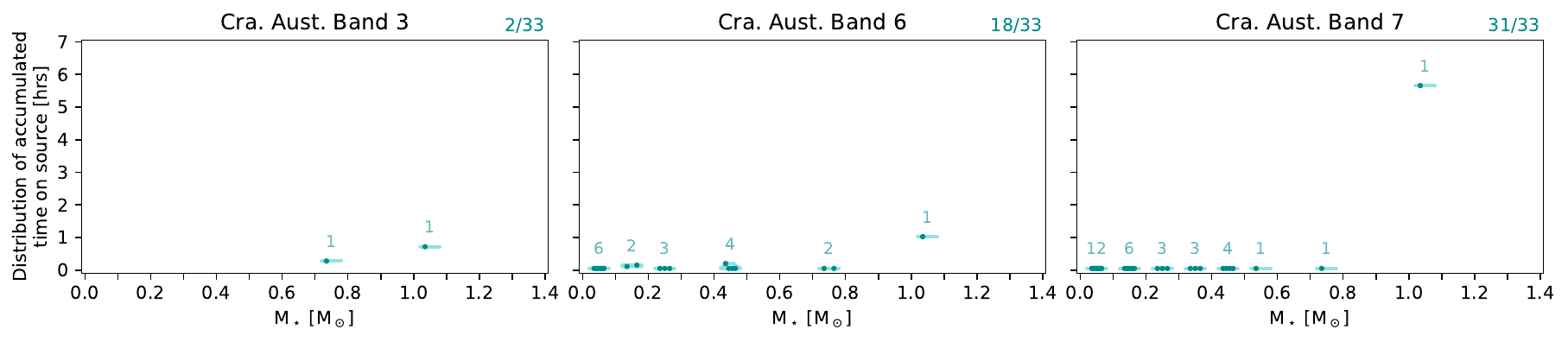}
    \includegraphics[width=0.9\textwidth]{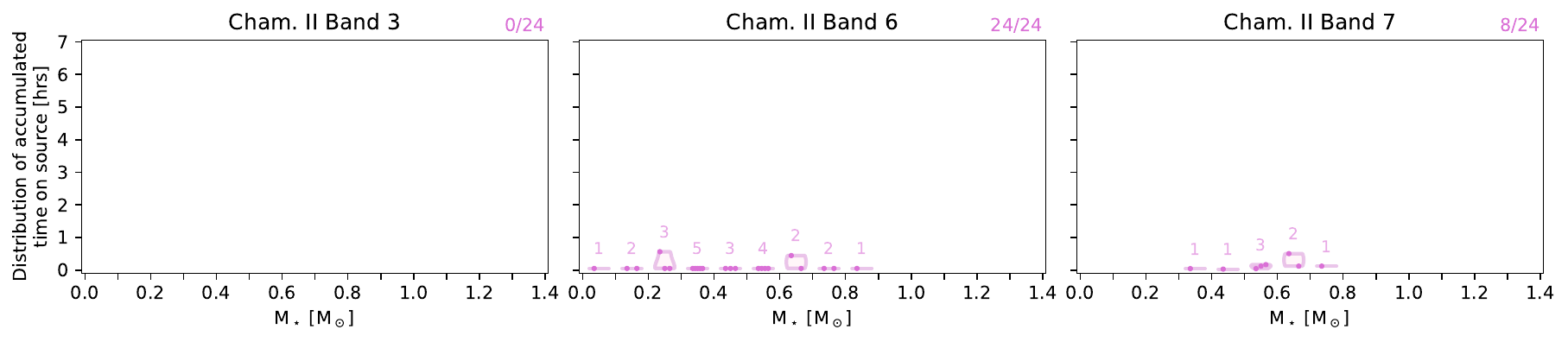}
    \caption{Distribution of accumulated exposure time per stellar mass bin, as in Fig.~\ref{fig:regions-time-per-band}. Only sources with known stellar masses are included in this figure. On the top right of each panel we show the fraction of sources with known stellar masses that have been observed in each band. The numbers above the distributions indicate the number of sources within that bin.  }
    \label{fig:app:regions-time-per-band}
\end{figure*}

\begin{figure*}
    \centering
    \includegraphics[width=0.8\textwidth]{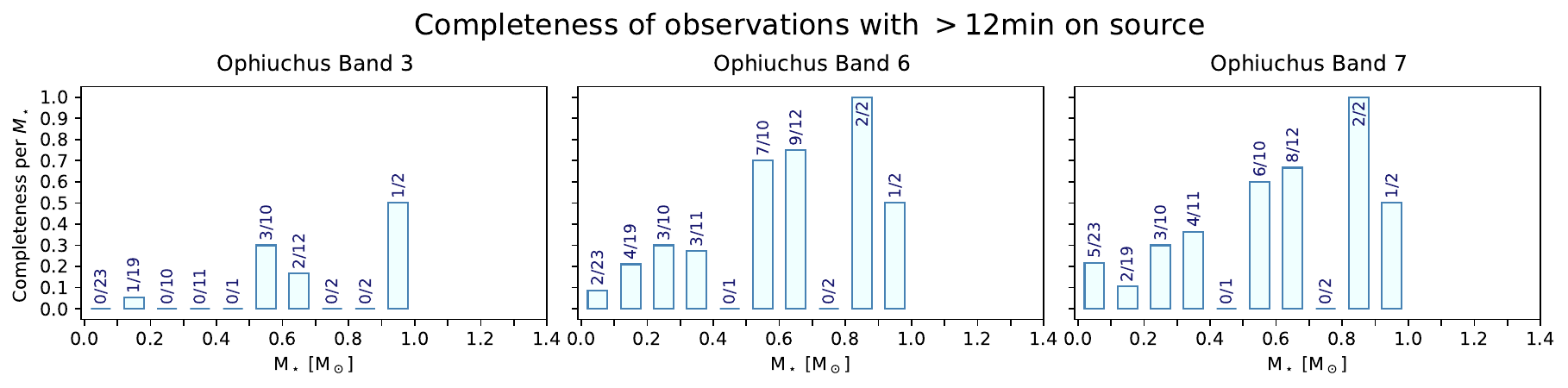}
    \includegraphics[width=0.8\textwidth]{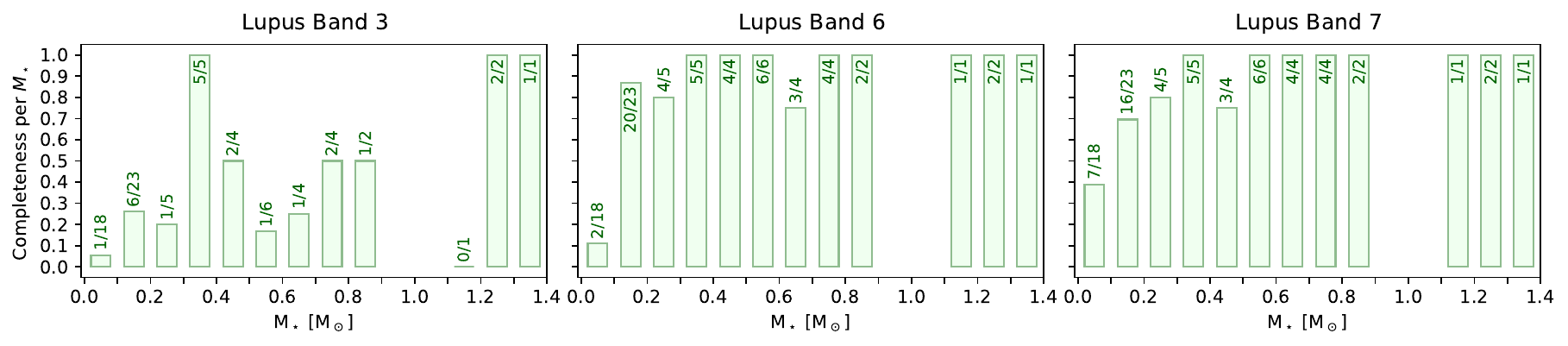}
    \includegraphics[width=0.8\textwidth]{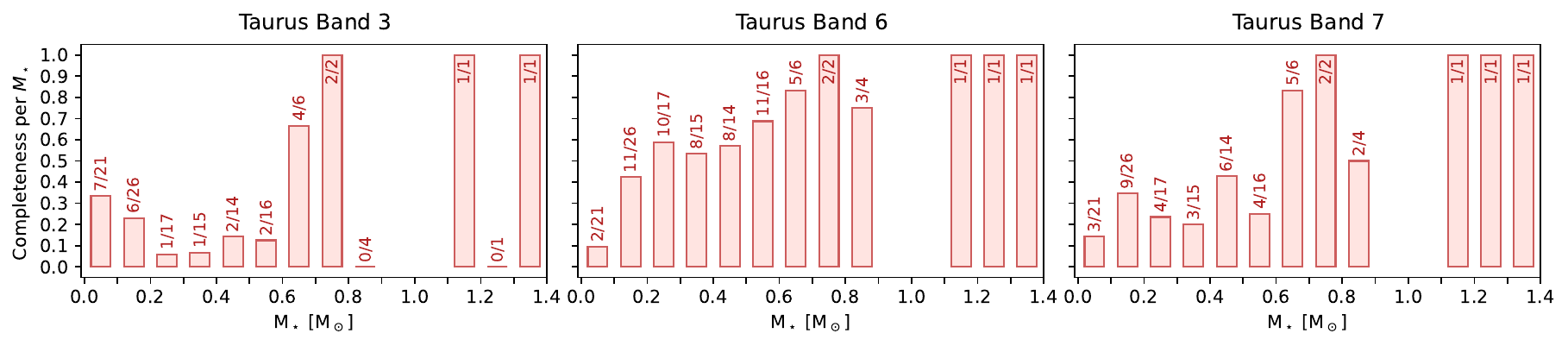}
    \includegraphics[width=0.8\textwidth]{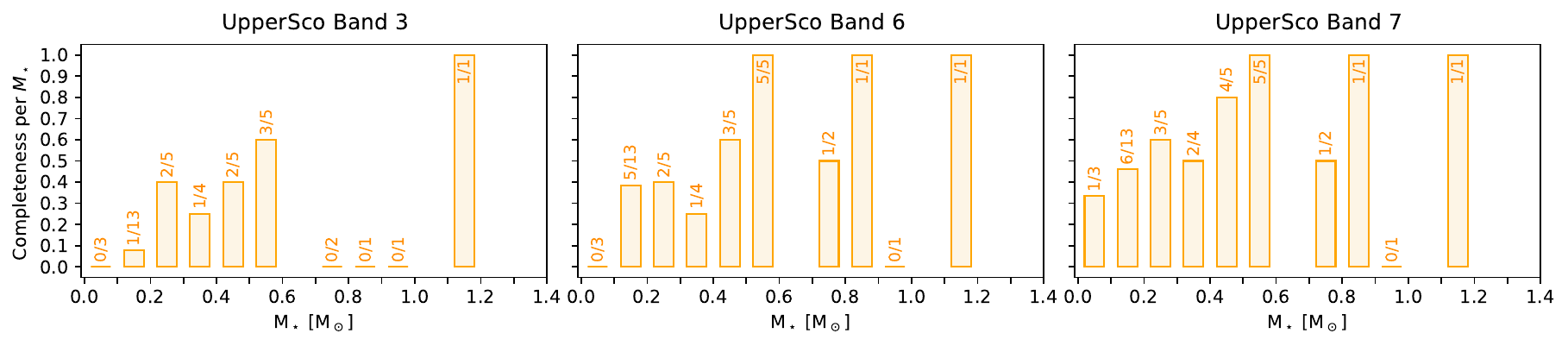}
    \includegraphics[width=0.8\textwidth]{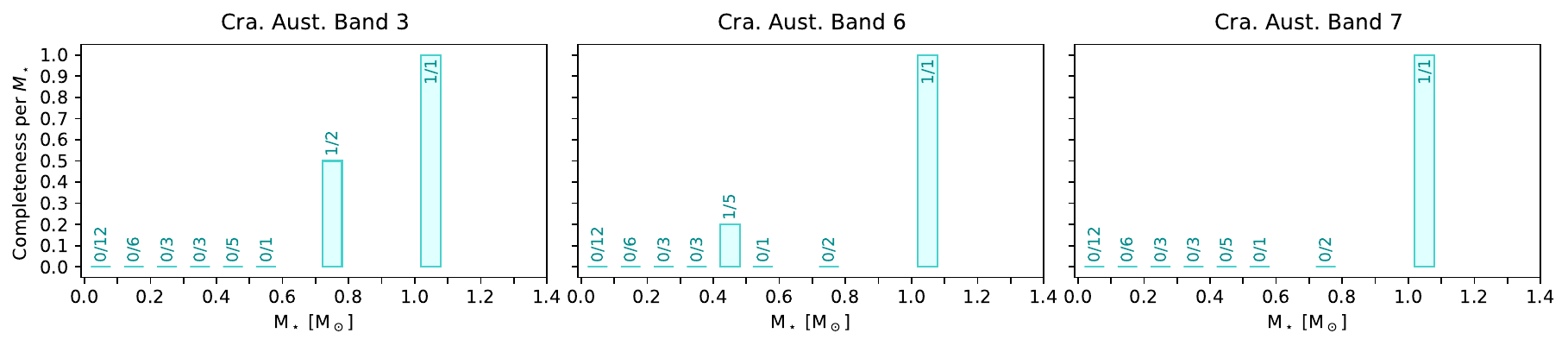}
    \includegraphics[width=0.8\textwidth]{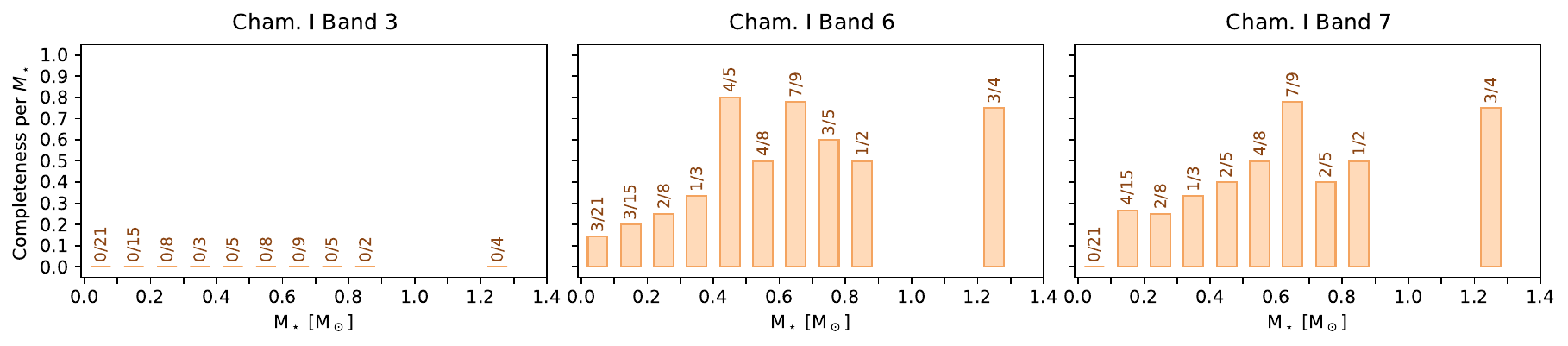}
    \includegraphics[width=0.8\textwidth]{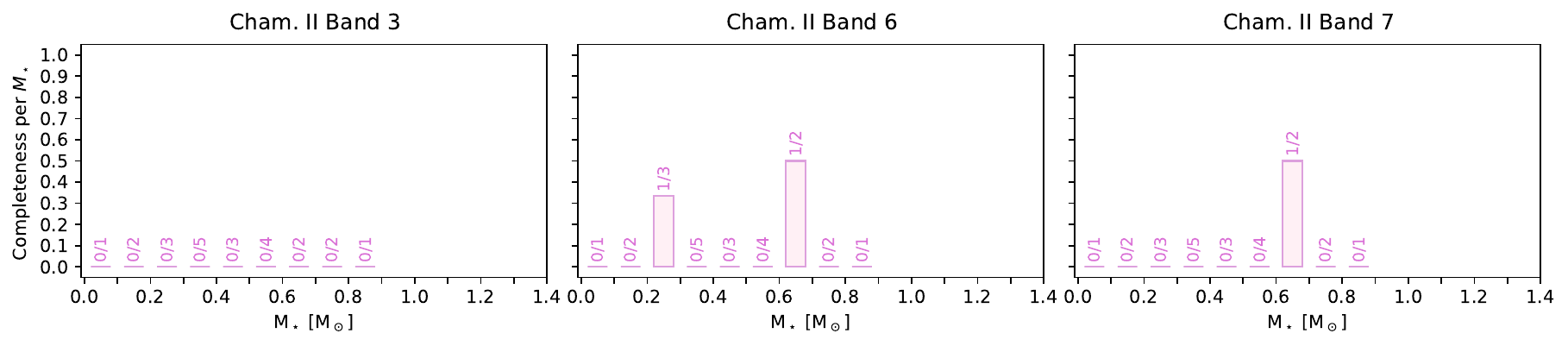}
    \caption{Completeness of observations above 12\,min on source, per stellar mass bin. The fraction above each bar shows the number of objects with more than 12\,min compared to the total number of objects in that mass range. }
    \label{fig:app:regions-completeness-per-band}
\end{figure*}

We include additional figures analyzing the observations of the ``Disks and planet formation'' category. 
\begin{itemize}
    \item In Fig.~\ref{fig:app:galactic_plane}, we present the observed coordinates in the galactic plane up to 500\,pc, adding the missing dimension of Fig.~\ref{fig:sky-distance-all}. TW\,Hya is shown with a star marker. 

    \item In Fig.~\ref{fig:app:pub-fraction}, we present the fraction of Project codes that have received at least $n$ publications, with $n$ ranging from 1 to 10. The number of repeated publications continues to increase for earlier cycles.
    
    \item In Fig.~\ref{fig:app:pie_bands-source}, we show the percentage of coordinates with observations in $n$ different ALMA Bands. For example, $46.2\%$ of coordinates have observations in only one band in Ophiuchus, $10.5\%$ have observations in two bands, $9.4\%$ have observations in three bands. Among the most observed nearby star-forming regions, Upper Sco has the lowest fraction of disks observed in at least two ALMA Bands. 
        
    \item In Fig.~\ref{fig:app:hist-time-per-ra}, we show the distribution of time on source per Right Ascension (RA), as well as the fraction of time spent on each scientific category per RA. In order to calculate regions of RA where observations are most frequent, we estimate a ``RA stress'', calculated by subtracting the average time per RA, normalizing it to the maximum residual, and convolving with a Gaussian filter with a full width at half maximum of 1\,hr. For reference, if all RA received the same amount of time, the RA stress would be constant and equal to zero. In this calculation, we have ignored the Declination of the coordinates, which will have an impact in the time that a coordinate is observable in the sky. We note that the integration time of mosaic observations is an average over the area of the observed region, computed from the total time on source multiplied by the ratio between the size of the primary beam and the observed area \citep{alma_archive_manual_cycle13}. In the case of the ``Disks and Planet formation'' category, the corrected time of mosaics adds about 40\,hrs to the total on source time. In Fig.~\ref{fig:app:hist-time-per-ra}, we have excluded the observations of the ``Solar System'' category. 

    The ``Disks and planet formation'' scientific category has its time concentrated at the RA of the nearby star-forming regions, a RA range that is also shared with the ``ISM and star formation''. The TWA region, containing TW\,Hya, is located at a RA with a RA stress of 0. The regions with the lowest observational stress are between 7 to 9\,hrs, and from 19 to 22\,hrs. 
    
    \item In Fig.~\ref{fig:app:regions-time-per-band}, we show a continuation of Fig.~\ref{fig:regions-time-per-band}, for the star-forming regions of Corona Australis and Chameleon II. 

    \item In Fig.~\ref{fig:app:regions-completeness-per-band}, we present the completeness fraction of observations longer than 12\,min for each stellar mass bin. We note that the stellar mass intrinsic uncertainty is not included in this figure, and therefore it should only be taken as a reference for completeness over different stellar populations. 
\end{itemize}

\section{C. Dust continuum galleries.}\label{sec:app:galleries}

\begin{figure*}[h!]
    \centering
    \includegraphics[width=0.96\textwidth]{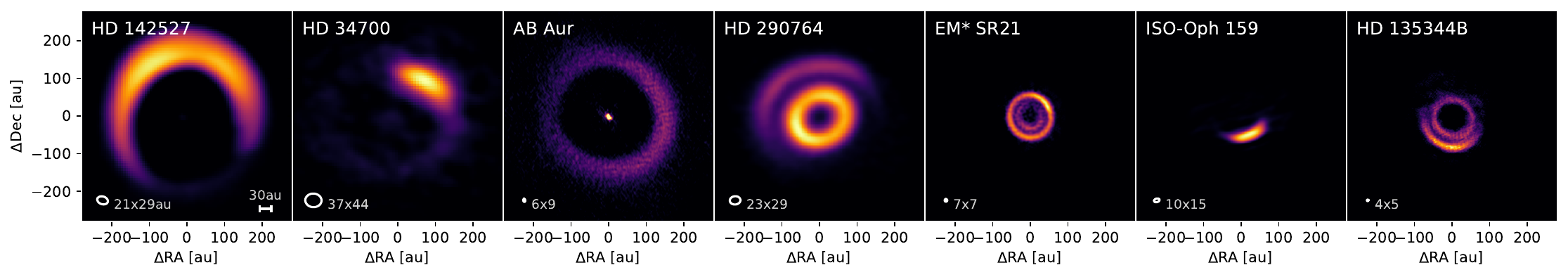}
    \caption{ Collection of disks with wide high-contrast asymmetries in the dust continuum emission, shown at the same spatial scale based on the Gaia DR3 parallax. The colorscale is chosen arbitrarily to highlight the substructures. These images were presented in \citet{yamaguchi2019} (HD\,142527), \citet{stadler2026} (HD\,34700), \citet{vioque2026} (AB Aur, HD\,290764), \citet{yang-yi2023} (EM$^*$\,SR21), \citet{yang-haifeng2023} (ISO-Oph\,159, or IRS\,48), and \citet{casassus2021} (HD\,135344B). }
    \label{fig:disks_large_asym}
\end{figure*}

\begin{figure*}[h!]
    \centering
    \includegraphics[width=0.96\textwidth]{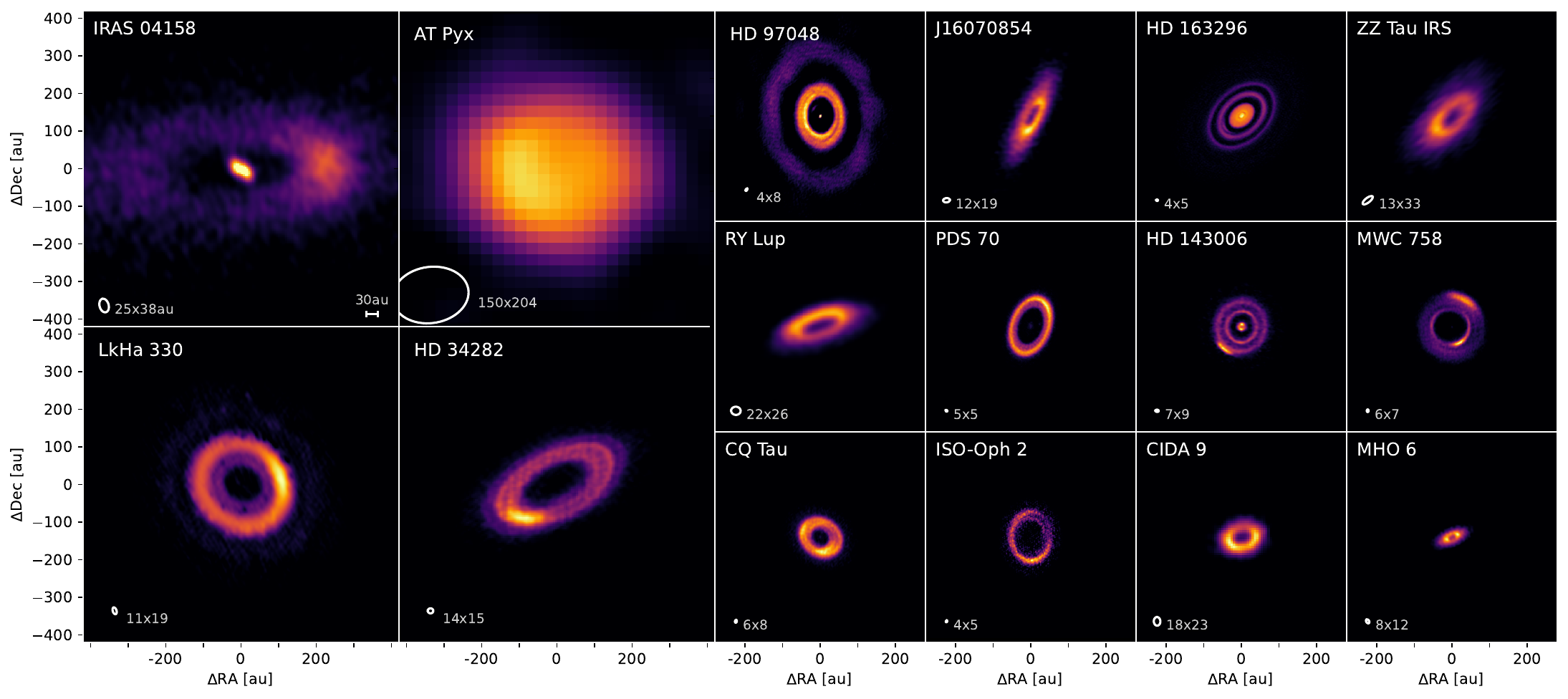}
    \caption{ Collection of disks with localized asymmetries in the dust continuum emission, shown at the same spatial scale based on the Gaia DR3 parallax. The colorscale is chosen arbitrarily to highlight the substructures. These images were presented in \citet{ragusa2021} (IRAS\,04158+2805), \citet{mclachlan2026} (AT\,Pyx), \citet{pinilla2022a} (LkHa\,330), \citet{francis2020} (HD\,34282), \citet{vioque2026} (HD\,97048, J16070854), \citet{isella2018} (HD\,163296), \citet{hashimoto2021b} (ZZ\,Tau\,IRS), \citet{norfolk2021} (RY\,Lup), \citet{benisty2021} (PDS70), \citet{perez2018} (HD\,143006), \citet{dong2018} (MWC\,758), \citet{zagaria2025} (CQ\,Tau), \citet{gonzalez-ruilova2020} (ISO-Oph\,2), \citet{long2018b} (CIDA\,9), and \citet{kurtovic2021} (MHO\,6). }
    \label{fig:disks_localized_asym}
\end{figure*}

\begin{figure*}[h!]
    \centering
    \includegraphics[width=0.96\textwidth]{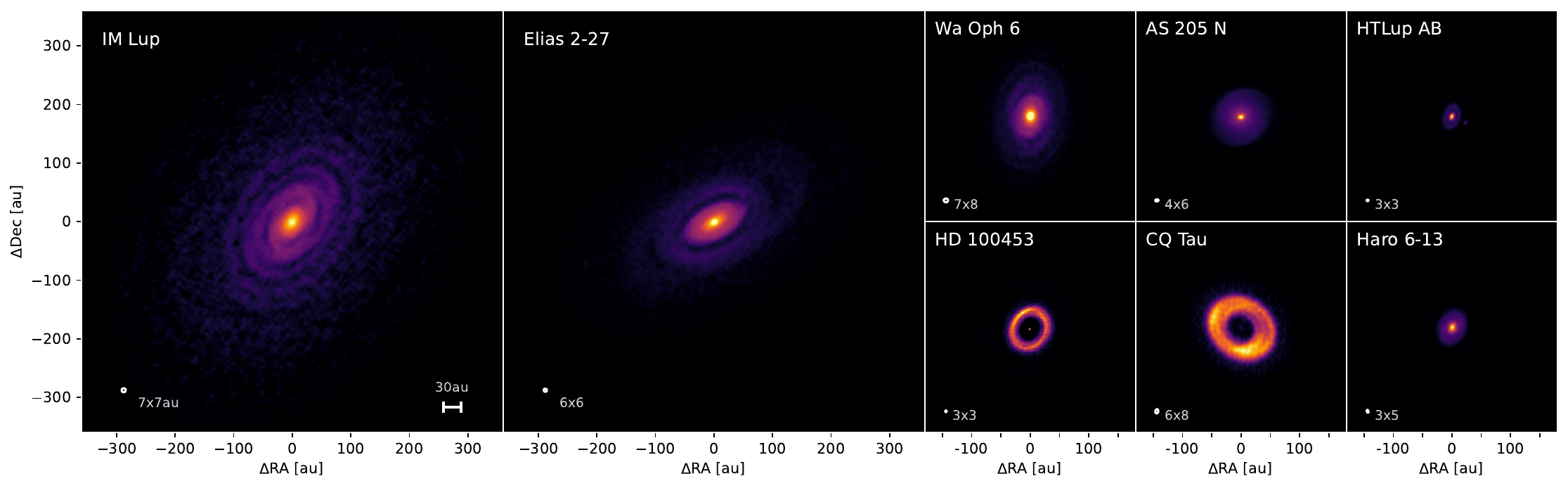}
    \caption{ Collection of disks with spirals in the dust continuum emission, shown at the same spatial scale based on the Gaia DR3 parallax. The colorscale is chosen arbitrarily to highlight the substructures. These images were presented in \citet{huang2018c} (IM\,Lup, Elias\,2-27, Wa\,Oph\,6), \citet{kurtovic2018} (AS\,205, HT\,Lup), \citet{rosotti2020} (HD\,100453), \citet{zagaria2025} (CQ\,Tau), and \citet{huang2025} (Haro\,6-13).}
    \label{fig:disks_spirals}
\end{figure*}

In Fig.~\ref{fig:disks_large_asym}, \ref{fig:disks_localized_asym}, and \ref{fig:disks_spirals}, we show a gallery of disks with non-axisymmetric dust continuum emission. All these disks are shown at the same spatial scale. The structures are split by high contrast crescents, asymmetries, or spirals, but we note that projection effects and continuum optical depths could contribute to hide or enhance the continuum contrast of a substructure \citep[see discussion by ][]{flores-rivera2026}. 

\begin{figure*}[h!]
    \centering
    \includegraphics[height=0.94\textheight]{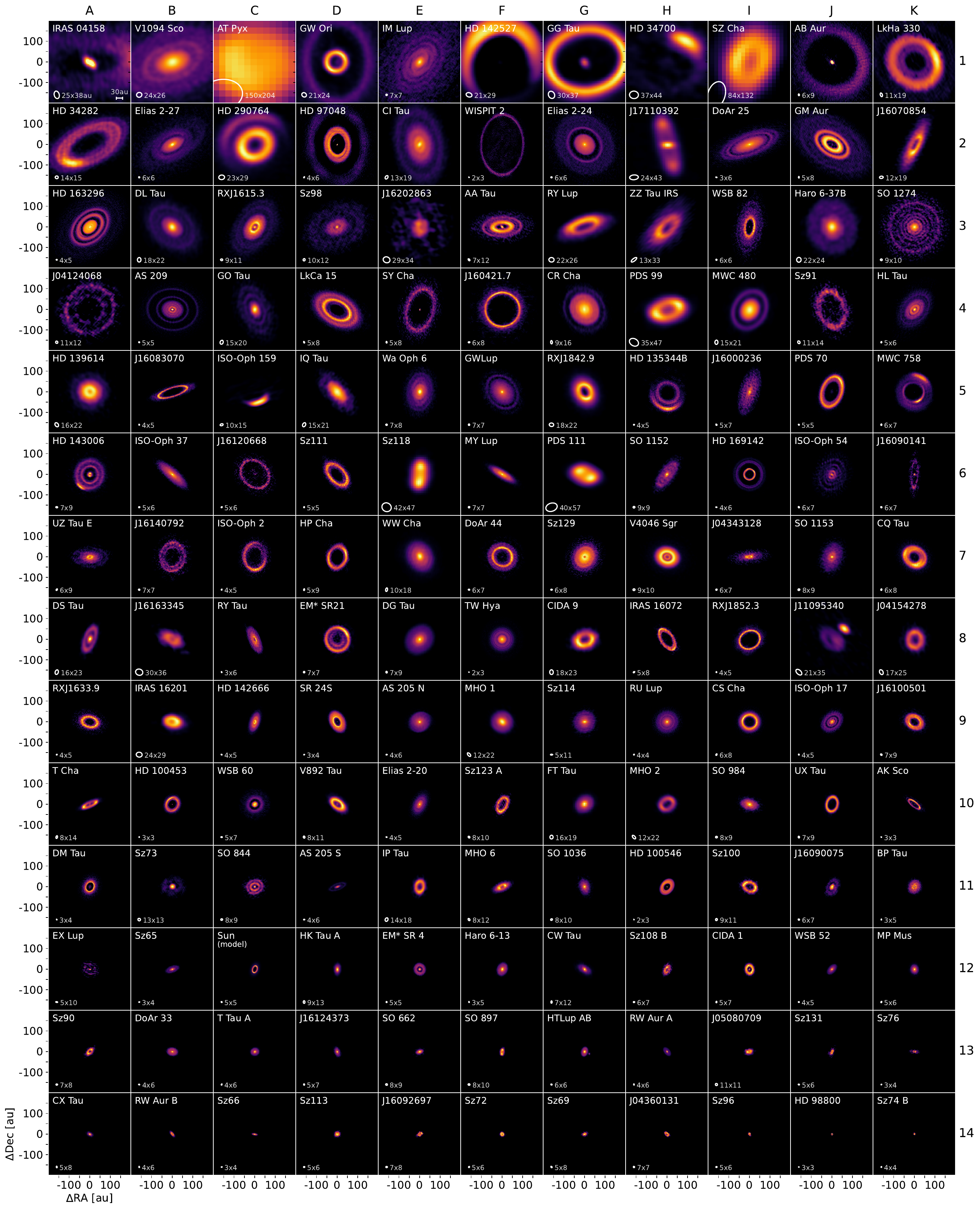}
    \caption{ Collection of disks with substructures in dust continuum emission, available in the literature. Each panel is 400\,au in size, and the disks are scaled based on the Gaia DR3 parallax, ranging from largest to smallest continuum disk from top to bottom. Further details and references in Sect.~\ref{sec:app:galleries}. }
    \label{fig:disks_rings}
\end{figure*}

In Fig.~\ref{fig:disks_rings}, we present a gallery of spatially resolved disks with substructures, also shown at the same spatial scale. In this gallery, we have given preference to Class II disks, mostly in nearby star-forming regions. The color scale was chosen arbitrarily for each disk to emphasize its substructures. We note that the continuum images in this gallery are predominantly from ALMA Band 6 or Band 7, but some of these disks are also shown in Bands 3, 4, or 8. All these images have at least one previous publication in the literature, listed in Tab.~\ref{tab:references_disks_1} and \ref{tab:references_disks_2}. Note the alphanumeric grid of the figure, to aid in the matching between the gallery and the tables.

\clearpage

\clearpage

\clearpage

\begin{sidewaystable}[h!]
    \centering
    \caption{Related references to disks in Figure \ref{fig:disks_rings}.}
    \begin{tabular}{llcl||llcl}
Name & Related   & Panel                      & Project & Name & Related   & Panel                      & Project \\
     & Reference & Fig.~\ref{fig:disks_rings} & code    &      & Reference & Fig.~\ref{fig:disks_rings} & code    \\
\hline\hline
IRAS 04158+2805 & \citet{ragusa2021} & A1 & 2019.1.00847.S & HD 139614 & \citet{bosschaart2026} & A5 & 2022.1.01302.S \\
\href{https://ke-coco-zhang.github.io/agepro-website/}{V1094 Sco} & \citet{deng2025} & B1 & 2021.1.00128.L & 2MASS J16083070-3828268 & \citet{guerra-alvarado2025} & B5 & 2018.1.00689.S \\
AT Pyx & \citet{mclachlan2026} & C1 & 2021.1.01705.S & ISO-Oph 159 & \citet{haifeng2023} & C5 & 2019.1.01059.S \\
GW Ori & \citet{vioque2026} & D1 & 2021.1.01661.S & IQ Tau & \citet{long2018b} & D5 & 2016.1.01164.S \\
\href{https://almascience.eso.org/almadata/lp/DSHARP/}{IM Lup} & \citet{huang2018c} & E1 & 2016.1.00484.L & \href{https://almascience.eso.org/almadata/lp/DSHARP/}{WaOph 6} & \citet{huang2018c} & E5 & 2016.1.00484.L \\
HD 142527 & \citet{yamaguchi2019} & F1 & 2012.1.00631.S & \href{https://almascience.eso.org/almadata/lp/DSHARP/}{GW Lup} & \citet{huang2018b} & F5 & 2016.1.00484.L \\
GG Tau & \citet{rota2024} & G1 & 2018.1.00532.S & \href{https://www.exoalma.com/data}{RX J1842.9-3532} & \citet{curone2025} & G5 & 2021.1.01123.L \\
HD 34700 & \citet{stadler2026} & H1 & 2022.1.00760.S & HD 135344B & \citet{casassus2021} & H5 & 2018.1.01066.S \\
SZ Cha & \citet{pascucci2016} & I1 & 2013.1.00437.S & 2MASS J16000236-4222145 & \citet{guerra-alvarado2025} & I5 & 2018.1.01054.S \\
AB Aur & \citet{vioque2026} & J1 & 2022.1.00313.S & PDS 70 & \citet{benisty2021} & J5 & 2018.A.00030.S \\
\href{http://cdsarc.u-strasbg.fr/viz-bin/cat/J/A+A/665/A128}{LkHa 330} & \citet{pinilla2022a} & K1 & 2018.1.01302.S & MWC 758 & \citet{dong2018} & K5 & 2017.1.00492.S \\
HD 34282 & \citet{francis2020} & A2 & 2017.1.01578.S & \href{https://almascience.eso.org/almadata/lp/DSHARP/}{HD 143006} & \citet{perez2018} & A6 & 2016.1.00484.L \\
\href{https://almascience.eso.org/almadata/lp/DSHARP/}{Elias 2-27} & \citet{huang2018c} & B2 & 2016.1.00484.L & ISO-Oph 37 & \citet{cieza2021} & B6 & 2018.1.00028.S \\
HD 290764 & \citet{woelfer2023} & C2 & 2016.1.01344.S & 2MASS J16120668-3010270 & \citet{jialin2025} & C6 & 2022.1.00646.S \\
HD 97048 & \citet{vioque2026} & D2 & 2019.1.01091.S & Sz 111 & \citet{guerra-alvarado2025} & D6 & 2018.1.00689.S \\
CI Tau & \citet{rosotti2021} & E2 & 2017.A.00014.S & Sz 118 & \citet{vioque2026} & E6 & 2022.1.00340.S \\
WISPIT 2 & \citet{facchini2026} & F2 & 2024.A.00064.S & \href{https://almascience.eso.org/almadata/lp/DSHARP/}{MY Lup} & \citet{huang2018b} & F6 & 2016.1.00484.L \\
\href{https://almascience.eso.org/almadata/lp/DSHARP/}{Elias 2-24} & \citet{huang2018b} & G2 & 2016.1.00484.L & \href{https://zenodo.org/records/14164743}{PDS 111} & \citet{derkink2024} & G6 & 2021.1.01705.S \\
2MASS J17110392-2722551 & \citet{alves2020} & H2 & 2013.1.00291.S & \href{https://zenodo.org/records/13821034}{SO 1152} & \citet{huang2024} & H6 & 2022.1.00728.S \\
\href{https://almascience.eso.org/almadata/lp/DSHARP/}{DoAr 25} & \citet{huang2018b} & I2 & 2016.1.00484.L & HD 169142 & \citet{perez2019} & I6 & 2016.1.00344.S \\
\href{https://zenodo.org/records/3628656}{GM Aur} & \citet{huang2020} & J2 & 2017.1.01151.S & \href{https://zenodo.org/records/17965437}{ISO-Oph 54} & \citet{jiang2026} & J6 & 2018.1.00028.S \\
2MASS J16070854-3914075 & \citet{vioque2026} & K2 & 2022.1.00742.S & 2MASS J16090141-3925119 & \citet{guerra-alvarado2025} & K6 & 2018.1.01054.S \\
\href{https://almascience.eso.org/almadata/lp/DSHARP/}{HD 163296} & \citet{isella2018} & A3 & 2016.1.00484.L & UZ Tau & \citet{vioque2026} & A7 & 2017.1.00388.S \\
DL Tau & \citet{long2018b} & B3 & 2016.1.01164.S & 2MASS J16140792-1938292 & \citet{vioque2026} & B7 & 2022.1.00646.S \\
RX J1615.3-3255 & \citet{vioque2026} & C3 & 2016.1.01286.S & ISO-Oph 2 & \citet{gonzalez-ruilova2020} & C7 & 2018.1.00028.S \\
Sz 98 & \citet{guerra-alvarado2025} & D3 & 2018.1.01458.S & HP Cha & \citet{norfolk2021} & D7 & 2017.1.01460.S \\
\href{https://ke-coco-zhang.github.io/agepro-website/}{2MASS J16202863-2442087} & \citet{miley2025} & E3 & 2021.1.00128.L & WW Cha & \citet{kanagawa2021} & E7 & 2017.1.00286.S \\
AA Tau & \citet{yamaguchi2024} & F3 & 2018.1.01829.S & DoAr 44 & \citet{cieza2021} & F7 & 2018.1.00028.S \\
RY Lup & \citet{norfolk2021} & G3 & 2017.1.00449.S & \href{https://almascience.eso.org/almadata/lp/DSHARP/}{Sz 129} & \citet{huang2018b} & G7 & 2016.1.00484.L \\
ZZ Tau IRS & \citet{hashimoto2021b} & H3 & 2016.1.01511.S & \href{https://www.exoalma.com/data}{V4046 Sgr} & \citet{curone2025} & H7 & 2021.1.01123.L \\
WSB 82 & \citet{cieza2021} & I3 & 2018.1.00028.S & 2MASS J04343128+1722201 & \citet{shi2024} & I7 & 2019.1.00566.S \\
Haro 6-37 B & \citet{bosschaart2026} & J3 & 2022.1.01302.S & \href{https://zenodo.org/records/13821034}{SO 1153} & \citet{huang2024} & J7 & 2022.1.00728.S \\
\href{https://zenodo.org/records/13821034}{SO 1274} & \citet{huang2024} & K3 & 2022.1.00728.S & CQ Tau & \citet{zagaria2025} & K7 & 2017.1.01404.S \\
2MASS J04124068+2438157 & \citet{long2023} & A4 & 2019.1.00566.S & DS Tau & \citet{long2018b} & A8 & 2016.1.01164.S \\
\href{https://almascience.eso.org/almadata/lp/DSHARP/}{AS 209} & \citet{guzman2018} & B4 & 2016.1.00484.L & \href{https://ke-coco-zhang.github.io/agepro-website/}{2MASS J16163345-2521505} & \citet{agurto-gangas2025} & B8 & 2021.1.00128.L \\
GO Tau & \citet{long2018b} & C4 & 2016.1.01164.S & WSB 60 & \citet{cieza2021} & C8 & 2018.1.00028.S \\
LkCa 15 & \citet{long2022} & D4 & 2018.1.00350.S & RY Tau & \citet{ribas2024} & D8 & 2017.1.01460.S \\
SY Cha & \citet{orihara2023} & E4 & 2018.1.00689.S & \href{https://zenodo.org/records/7760435}{EM* SR 21} & \citet{yang-yi2023} & E8 & 2018.1.00689.S \\
2MASS J16042165-2130284 & \citet{stadler2023} & F4 & 2018.1.01255.S & DG Tau & \citet{ohashi2023b} & F8 & 2015.1.01268.S \\
CR Cha & \citet{seongjoong2020} & G4 & 2017.1.00286.S & \href{https://dataverse.harvard.edu/dataset.xhtml?persistentId=doi:10.7910/DVN/PXDKBC}{TW Hya} & \citet{huang2018a} & G8 & 2015.A.00005.S \\
PDS 99 & \citet{francis2020} & H4 & 2015.1.01301.S & CIDA 9 & \citet{harsono2024} & H8 & 2016.1.01164.S \\
\href{https://alma-maps.info/data.html}{MWC 480} & \citet{sierra2021} & I4 & 2018.1.01055.L & IRAS 16072-2057 & \citet{vioque2026} & I8 & 2022.1.00646.S \\
Sz 91 & \citet{mauco2021} & J4 & 2018.1.01020.S & RX J1852.3-3700 & \citet{vioque2026} & J8 & 2018.1.00689.S \\
\href{https://almascience.eso.org/alma-data/science-verification/overview}{HL Tau} & \citet{carrasco-gonzalez2019} & K4 & 2011.0.00015.SV & 2MASS J11095340-7634255 & \citet{bosschaart2026} & K8 & 2022.1.01302.S \\
\hline\hline
    \end{tabular}
      \begin{minipage}{16cm}
        \vspace{0.1cm}
        Note: Only one related reference is listed for each image, with preference to the publication that made the image publicly available (see links in source name), the first publication of the image, or the first in depth analysis. More publications may be associated with these images. When available, public repositories are linked to the name of the source. 
      \end{minipage}
    \label{tab:references_disks_1}
\end{sidewaystable}

\begin{sidewaystable}[h!]
    \centering
    \caption{Related references to disks in Figure \ref{fig:disks_rings} - continuation.}
    \begin{tabular}{llcl||llcl}
Name & Related   & Panel                      & Project & Name & Related   & Panel                      & Project \\
     & Reference & Fig.~\ref{fig:disks_rings} & code    &      & Reference & Fig.~\ref{fig:disks_rings} & code    \\
\hline\hline
2MASS J04154278+2909597 & \citet{bosschaart2026} & A9 & 2022.1.01302.S & EX Lup & \citet{yamaguchi2025} & A12 & 2017.1.00388.S \\
RX J1633.9-2442 & \citet{cieza2021} & B9 & 2018.1.00028.S & \href{https://cdsarc.cds.unistra.fr/viz-bin/cat/J/A+A/682/A55#/article}{Sz 65} & \citet{miley2024} & B12 & 2018.1.00271.S \\
IRAS 16201-2410 & \citet{dasgupta2025} & C9 & 2021.1.00378.S & \href{https://simio-continuum.readthedocs.io/}{Sun} & \citet{bergez-casalou2022} & C12 & - \\
\href{https://almascience.eso.org/almadata/lp/DSHARP/}{HD 142666} & \citet{huang2018b} & D9 & 2016.1.00484.L & \href{https://cdsarc.cds.unistra.fr/viz-bin/cat/J/A+A/642/A164#/browse}{HK Tau A} & \citet{villenave2020} & D12 & 2016.1.00460.S \\
EM* SR 24 S & \citet{cieza2021} & E9 & 2018.1.00028.S & \href{https://almascience.eso.org/almadata/lp/DSHARP/}{EM* SR 4} & \citet{huang2018b} & E12 & 2016.1.00484.L \\
\href{https://almascience.eso.org/almadata/lp/DSHARP/}{AS 205 N} & \citet{kurtovic2018} & F9 & 2016.1.00484.L & \href{https://zenodo.org/records/15002753}{Haro 6-13} & \citet{huang2025} & F12 & 2022.1.01365.S \\
MHO 1 & \citet{bosschaart2026} & G9 & 2022.1.01302.S & CW Tau & \citet{ueda2022} & G12 & 2019.1.01108.S \\
\href{https://almascience.eso.org/almadata/lp/DSHARP/}{Sz 114} & \citet{huang2018b} & H9 & 2016.1.00484.L & Sz 108 B & \citet{guerra-alvarado2025} & H12 & 2022.1.00154.S \\
\href{https://almascience.eso.org/almadata/lp/DSHARP/}{RU Lup} & \citet{huang2018b} & I9 & 2016.1.00484.L & CIDA 1 & \citet{pinilla2021} & I12 & 2018.1.00536.S \\
\href{https://github.com/nicokurtovic/CSCha_ALMA_2017.1.00969.S}{CS Cha} & \citet{kurtovic2022} & J9 & 2017.1.00969.S & \href{https://almascience.eso.org/almadata/lp/DSHARP/}{WSB 52} & \citet{huang2018b} & J12 & 2016.1.00484.L \\
ISO-Oph 17 & \citet{cieza2021} & K9 & 2018.1.00028.S & MP Mus & \citet{ribas2025} & K12 & 2022.1.01758.S \\
\href{http://cdsarc.u-strasbg.fr/viz-bin/cat/J/A+A/639/A121}{J16100501} & \citet{facchini2023} & A10 & 2018.1.01255.S & Sz 90 & \citet{guerra-alvarado2025} & A13 & 2022.1.00154.S \\
T Cha & \citet{hendler2018} & B10 & 2015.1.00979.S & \href{https://almascience.eso.org/almadata/lp/DSHARP/}{DoAr 33} & \citet{huang2018b} & B13 & 2016.1.00484.L \\
HD 100453 & \citet{rosotti2020} & C10 & 2017.1.01678.S & T Tau & \citet{stapper2025} & C13 & 2019.1.00703.S \\
V892 Tau & \citet{long2021} & D10 & 2021.1.01137.S & 2MASS J16124373-3815031 & \citet{guerra-alvarado2025} & D13 & 2022.1.00154.S \\
\href{https://almascience.eso.org/almadata/lp/DSHARP/}{Elias 2-20} & \citet{huang2018b} & E10 & 2016.1.00484.L & \href{https://zenodo.org/records/13821034}{SO 662} & \citet{huang2024} & E13 & 2022.1.00728.S \\
Sz 123 A & \citet{guerra-alvarado2025} & F10 & 2018.1.01458.S & \href{https://zenodo.org/records/13821034}{SO 897} & \citet{huang2024} & F13 & 2022.1.00728.S \\
FT Tau & \citet{long2018b} & G10 & 2016.1.01164.S & \href{https://almascience.eso.org/almadata/lp/DSHARP/}{HT Lup A} & \citet{kurtovic2018} & G13 & 2016.1.00484.L \\
MHO 2 & \citet{bosschaart2026} & H10 & 2022.1.01302.S & \href{https://zenodo.org/records/12825068}{RW Aur A} & \citet{kurtovic2024c} & H13 & 2017.1.01631.S \\
\href{https://zenodo.org/records/13821034}{SO 984} & \citet{huang2024} & I10 & 2022.1.00728.S & 2MASS J05080709+2427123 & \citet{shi2024} & I13 & 2019.1.00566.S \\
UX Tau & \citet{vioque2026} & J10 & 2021.1.00994.S & Sz 131 & \citet{guerra-alvarado2025} & J13 & 2022.1.00154.S \\
AK Sco & \citet{vioque2026} & K10 & 2019.1.01210.S & Sz 76 & \citet{guerra-alvarado2025} & K13 & 2018.1.01054.S \\
DM Tau & \citet{hashimoto2021a} & A11 & 2018.1.01755.S & CX Tau & \citet{facchini2019} & A14 & 2016.1.00715.S \\
Sz 73 & \citet{guerra-alvarado2025} & B11 & 2018.1.01458.S & \href{https://zenodo.org/records/12825068}{RW Aur B} & \citet{kurtovic2024c} & B14 & 2017.1.01631.S \\
\href{https://zenodo.org/records/13821034}{SO 844} & \citet{huang2024} & C11 & 2022.1.00728.S & \href{https://cdsarc.cds.unistra.fr/viz-bin/cat/J/A+A/682/A55#/article}{Sz 66} & \citet{miley2024} & C14 & 2018.1.00271.S \\
\href{https://almascience.eso.org/almadata/lp/DSHARP/}{AS 205 S} & \citet{kurtovic2018} & D11 & 2016.1.00484.L & Sz 113 & \citet{guerra-alvarado2025} & D14 & 2022.1.00154.S \\
IP Tau & \citet{long2018b} & E11 & 2016.1.01164.S & 2MASS J16092697-3836269 & \citet{guerra-alvarado2025} & E14 & 2022.1.00154.S \\
\href{https://github.com/nicokurtovic/VLMS_ALMA_2018.1.00310.S}{MHO 6} & \citet{kurtovic2021} & F11 & 2018.1.00310.S & Sz 72 & \citet{guerra-alvarado2025} & F14 & 2022.1.00154.S \\
\href{https://zenodo.org/records/13821034}{SO 1036} & \citet{huang2024} & G11 & 2022.1.00728.S & Sz 69 & \citet{guerra-alvarado2025} & G14 & 2018.1.01458.S \\
HD 100546 & \citet{casassus2022} & H11 & 2018.1.01309.S & 2MASS J04360131+1726120 & \citet{shi2024} & H14 & 2019.1.00566.S \\
Sz 100 & \citet{guerra-alvarado2025} & I11 & 2018.1.01054.S & Sz 96 & \citet{guerra-alvarado2025} & I14 & 2022.1.00154.S \\
2MASS J16090075-1908526 & \citet{vioque2025} & J11 & 2017.1.01167.S & HD 98800 & \citet{kennedy2019} & J14 & 2017.1.00350.S \\
BP Tau & \citet{gasman2025} & K11 & 2019.1.00607.S & Sz 74 B & \citet{guerra-alvarado2025} & K14 & 2022.1.00154.S \\
\hline\hline
    \end{tabular}
      \begin{minipage}{20cm}
        \vspace{0.1cm}
        Note: Only one related reference is listed for each image, with preference to the first publication or the first in depth analysis. More publications may be associated with these images. When available, public repositories are linked to the name of the source. For reference, a simulation of the disk around the Sun during its Class II phase is included in panel C12, original from \citet{bergez-casalou2022} and processed with the SIMIO package \citep{kurtovic2024a}. 
      \end{minipage}
    \label{tab:references_disks_2}
\end{sidewaystable}

\clearpage

\bibliographystyle{aasjournal}
\bibliography{ms}

\end{document}